\documentclass{article}
\usepackage[utf8]{inputenc}
\usepackage[english]{babel}
\usepackage[a4paper, margin=3.0cm]{geometry}
\usepackage{amsmath}
\usepackage{amssymb}
\usepackage{bbm}
\usepackage{braket}
\usepackage{caption}
\usepackage{pdfpages}
\usepackage{comment}
\usepackage{titlesec}
\usepackage{multirow}
\usepackage{adjustbox}
\usepackage{subcaption}
\usepackage{graphicx}
\usepackage{array}
\usepackage{hyperref}
\usepackage{url}
\usepackage{booktabs}
\usepackage{float}
\usepackage{cite}
\usepackage{adjustbox}
\usepackage[title]{appendix}

\usepackage{orcidlink}
\usepackage{graphicx}

\DeclareMathOperator{\Tr}{\mathrm{Tr}}
\DeclareMathOperator{\SU}{\mathrm{SU}}
\DeclareMathOperator{\Z}{\mathbb{Z}}
\newcommand{\real}{{\rm Re\,}}

\newcommand{\ide}{\mathbbm{1}}
\newcommand{\TTbar}{T\overline{T}}
\begin{document}

\begin{titlepage}
\begin{center}
{\Large\bf Confining strings in three-dimensional gauge theories beyond the Nambu--Got{\=o} approximation}
\end{center}
\vskip1.3cm
\centerline{
Michele~Caselle\orcidlink{0000-0001-5488-142X}\footnote{\href{mailto:caselle@to.infn.it}{{\tt caselle@to.infn.it}}}$^a$,
Nicodemo Magnoli\orcidlink{0000-0001-5372-4836}\footnote{\href{malito:magnoli@ge.infn.it}{\tt magnoli@ge.infn.it}}$^b$,
Alessandro~Nada\orcidlink{0000-0002-1766-5186}\footnote{\href{mailto:alessandro.nada@unito.it}{{\tt alessandro.nada@unito.it}}}$^a$,
}
\centerline{
Marco~Panero\orcidlink{0000-0001-9477-3749}\footnote{\href{mailto:marco.panero@unito.it}{{\tt marco.panero@unito.it}}}$^a$,
Dario~Panfalone\orcidlink{0009-0007-6651-7490}\footnote{\href{mailto:dario.panfalone@unito.it}{{\tt dario.panfalone@unito.it}}}$^a$ and
Lorenzo~Verzichelli\orcidlink{0009-0008-0825-4845}\footnote{\href{mailto:lorenzo.verzichelli@unito.it}{{\tt lorenzo.verzichelli@unito.it}}}$^a$
}
\vskip1.5cm
\centerline{$^a$ {\sl Department of Physics, University of Turin and INFN, Turin}}
\centerline{\sl Via Pietro Giuria 1, I-10125 Turin, Italy}
\vskip1.0cm

\centerline{$^b$ {\sl Department of Physics, University of Genoa and INFN, Genoa}}
\centerline{\sl Via Dodecaneso 33, I-16146, Genoa, Italy}
\vskip1.0cm

\setcounter{footnote}{0}
\renewcommand\thefootnote{\mbox{\arabic{footnote}}}
\begin{abstract}
\noindent We carry out a systematic study of the effective bosonic string describing confining flux tubes in $\SU(N)$ Yang--Mills theories in three spacetime dimensions. While their low-energy properties are known to be universal and are described well by the Nambu--Got{\=o} action, a non-trivial dependence on the gauge group is encoded in a series of undetermined subleading corrections in an expansion around the limit of an arbitrarily long string. We quantify the first two of these corrections by means of high-precision Monte Carlo simulations of Polyakov-loop correlators in the lattice regularization. We compare the results of novel lattice simulations for theories with $N=3$ and $6$ color charges, and report an improved estimate for the $N=2$ case, discussing the approach to the large-$N$ limit. Our results are compatible with analytical bounds derived from the S-matrix bootstrap approach. In addition, we also present a new test of the Svetitsky--Yaffe conjecture for the $\SU(3)$ theory in three dimensions, finding that the lattice results for the Polyakov-loop correlation function are in excellent agreement with the predictions of the Svetitsky--Yaffe mapping, which are worked out quantitatively applying conformal perturbation theory to the three-state Potts model in two dimensions. The implications of these results are discussed.
\end{abstract}

\end{titlepage}

\section{Introduction}

Effective string theory (EST) provides an accurate description of the long-wavelength dynamics of confining flux tubes in Yang--Mills theories; in this picture, confining flux tubes are modelled as thin, fluctuating bosonic strings~\cite{Luscher:1980ac, Luscher:1980fr}. In the past few decades, analytical and numerical studies of this description have provided invaluable insight in the understanding of confinement. In recent years a crucial feature of EST was found, that dramatically increased its predictiveness: namely, when looking at the large-distance expansion of the effective string action, the first few terms are \textit{universal} and correspond to the Nambu--Got{\=o} (NG) action~\cite{Nambu:1969se, Nambu:1974zg, Goto:1971ce}. This result goes under the name of ``low-energy universality'' and is a direct consequence of the symmetry constraints imposed by the Poincar\'e invariance in the target space~\cite{Luscher:2004ib, Meyer:2006qx, Aharony:2009gg, Aharony:2011gb, Gliozzi:2011hj, Gliozzi:2012cx, Dubovsky:2012sh, Aharony:2013ipa}. This universality is indeed observed in high-precision results from lattice simulations, as reviewed in refs.~\cite{Brandt:2016xsp, Caselle:2021eir}.

Thanks to its relative simplicity, the Nambu--Got{\=o} action has been studied for many decades: the main implications that can be derived analytically for this type of bosonic string have become standard textbook material in any introductory course on string theory~\cite{Polchinski:1998rq} (while those that cannot be studied analytically are now being addressed by machine-learning methods~\cite{Caselle:2023mvh}). By contrast, the determination of the terms beyond the Nambu--Got{\=o} action (BNG) in the effective string theory for confinement remains an open problem. Those terms appear only at high orders in an expansion around the limit of an infinitely long string, and thus prove somewhat elusive to study, but they encode crucial pieces of information for a proper characterization of the infrared dynamics of confining theories. As a matter of fact, even though the agreement between predictions of the Nambu--Got{\=o} model with Yang--Mills theories based on completely different gauge groups is striking, the signatures that characterize the differences between these gauge theories are necessarily encoded in the terms beyond the NG approximation. A quantitative investigation of the behavior of these terms in the EST would provide invaluable insight into the features of different confining gauge theories.

The most straightforward way to study the non-universal EST terms in numerical simulations on the lattice consists in investigating the two-point correlation function of static color sources, i.e., the Polyakov-loop correlator. As is well known, the presence of boundary terms in the EST action hinders the detection of bulk corrections at zero temperature, but it can be shown that the effects due to boundary terms become subleading and can be neglected at high temperatures~\cite{Caselle:2021eir}; thus, in the present work we will focus on the study of the Polyakov loop correlator at high temperatures in the confining phase of the gauge theory, i.e., for temperatures $T$ approaching the deconfinement transition from below.

In particular, we will study the fine details of the large-distance behavior of the Polyakov loop correlator of the $\SU(3)$ and $\SU(6)$ gauge theories in $D=2+1$ spacetime dimensions. The same strategy was followed in ref.~\cite{Caristo:2021tbk} for the $\SU(2)$ gauge theory: this work is meant as a natural continuation of this specific approach, and some results of that article will be elaborated further upon and comparatively discussed in our analysis. In passing, we also mention that another lattice study of the $\SU(3)$ theory in three dimensions, focused on the effective string picture for confining flux tubes at finite temperature, was reported in ref.~\cite{Bialas:2009pt}.

The choice of the number of color charges that we consider in the present study ($N=3$ and $N=6$) is not random: in $2+1$ dimensions the $\SU(3)$ theory has a second-order deconfinement phase transition, while in the case of $\SU(6)$ the transition is clearly of the first order.\footnote{Note that first-order deconfinement transitions are present also in the $\SU(4)$~\cite{Holland:2007ar} and $\SU(5)$ theories~\cite{Holland:2005nd}, albeit weak ones.} The quantitative and qualitative differences between $\SU(3)$ and $\SU(6)$ Yang--Mills theories are expected to induce characteristic signatures that should manifest themselves in the non-universal coefficients in the expansion of the EST action.

We also remark that the study of the $\SU(3)$ Yang--Mills theory in $2+1$ dimensions is relevant in the context of the Svetitsky--Yaffe conjecture~\cite{Svetitsky:1982gs}: since the deconfinement transition is continuous, in the proximity of the critical temperature one expects the low-energy features of the theory to be described by the two-dimensional spin model with global symmetry given by the center of the gauge group, i.e., $\Z_3$. In this case, we will examine the Polyakov-loop correlator \textit{quantitatively}, comparing it with the behavior predicted for the spin-spin correlator of the two-dimensional three-state Potts model.\footnote{Earlier studies of this subject include those reported in refs.~\cite{Christensen:1992is, Engels:1996dz}.} Since the Svetitsky--Yaffe conjecture may even have phenomenological implications for the phase diagram of quantum chromodynamics (QCD) and for relativistic nuclear collisions~\cite{Aarts:2023vsf}, it is important to test its validity and the numerical accuracy of its predictions in a setting where they can be compared with high-precision lattice results and for different universality classes, as is the case for $\SU(2)$ and $\SU(3)$ Yang--Mills theory in $2+1$ dimensions.

Finally, the investigation of $\SU(N)$ gauge theories in $2+1$ dimensions for increasing values of $N$ is of interest in view of the 't~Hooft limit, too~\cite{tHooft:1973alw}. Analytical studies specifically focused on these theories include estimates of the vacuum wavefunction, of the string tension, of the glueball spectrum, and of the properties in the high-temperature phase~\cite{Karabali:1995ps, Karabali:1997wk, Karabali:1998yq, Karabali:2000gy, Leigh:2005dg, Leigh:2006vg, Karabali:2009rg, Bicudo:2013yza, Bicudo:2014cra, Frasca:2016sky}; in parallel, these quantities have also been investigated in many numerical studies on the lattice~\cite{Teper:1998te, Johnson:2000qz, Lucini:2002wg, Meyer:2003wx, Meyer:2004jc, Bursa:2005tk, Bringoltz:2006zg, Athenodorou:2007du, Liddle:2008kk, Bringoltz:2008nd, Athenodorou:2008cj, Athenodorou:2011rx, Caselle:2011fy, Caselle:2011mn, Bursa:2012ab, Bialas:2012qz, Athenodorou:2013ioa, Athenodorou:2015nba, Lau:2015cna, Athenodorou:2016ebg, Athenodorou:2016kpd, Conkey:2019blu}.

The structure of this article is the following: in section~\ref{sec:est} we will review the details of the effective string model that are relevant for this work, with a particular focus on the predictions at high-temperature and beyond the NG approximation. In section~\ref{sec:lattice} we will briefly examine the technical details of the lattice-regularized gauge theories under study and of the numerical determination of the relevant observables; particular emphasis will be put on the procedure to identify the corrections beyond the NG approximation. Section~\ref{sec:svetitsky_yaffe} will be devoted to an in-depth description of the mapping between the $\SU(3)$ theory and the three-state Potts model. In section~\ref{sec:results} we will provide a complete analysis of the numerical results, first concerning the EST corrections to the Nambu--Got{\=o} and then on the Svetistky--Yaffe mapping. We will draw the main conclusions of this work in section~\ref{sec:conclusions}.

\section{Effective string theory of the color flux tube}
\label{sec:est}

Effective string theory relates the confinement of color sources to the formation of a thin string-like flux tube which leads, for large separations between the color charges, to the formation of a linearly confining potential. The partition function of the EST of choice is directly related to the correlator of two static color sources: in the lattice gauge theory setup that will be examined in detail in section~\ref{sec:lattice}, this corresponds to having a quantitative analytic description of the Polyakov loop correlator.

The simplest Poincar\'e-invariant EST is the well-known Nambu--Got{\=o} (NG) string model. In this picture, the string action $S_{\mathrm{NG}}$ is defined as follows:
\begin{align}
 S_{\mathrm{NG}}=\sigma_0 \int_{\Sigma} \mathrm{d}^2\xi \sqrt{g},
\end{align}
where $g \equiv \mathrm{det} g_{\alpha\beta}$ and 
\begin{align}
 g_{\alpha\beta} = \partial_\alpha X_\mu \partial_\beta X^\nu
\end{align}
is the metric induced on the reference world-sheet surface $\Sigma$; $\xi \equiv (\xi^0,\xi^1)$ denote the world-sheet coordinates. This term has a simple geometric interpretation: it measures the area of the surface spanned by the string in the target space and has only one free parameter, the string tension $\sigma_0$, of dimension two.

The Nambu--Got{\=o} action $S_{\mathrm{NG}}$ is clearly reparametrization-invariant and, in order to perform calculations, the first step is to fix this invariance. The standard choice is the ``physical gauge'', in which the two world-sheet coordinates are identified with the longitudinal degrees of freedom of the string: $\xi^0=X^0$ and $\xi^1=X^1$. In this case, the string action can be expressed as a function only of the $(D-2)$ degrees of freedom corresponding to the transverse displacements $X^i$, with $i=2,\dots,(D-1)$, which are assumed to be single-valued functions of the world-sheet coordinates. 

With this choice the determinant of the metric has the form
\begin{align}
g = 1 + \partial_0 X_i \partial_0 X^i + \partial_1 X_i \partial_1 X^i + \partial_0 X_i \partial_0 X^i \partial_1 X_j \partial_1 X^j - (\partial_0 X_i\partial_1 X_i)^2
\end{align}
and the Nambu--Got{\=o} action can then be expressed as a low-energy expansion in the number of derivatives of the transverse degrees of freedom of the string which, by an appropriate redefinition of the fields, can be rephrased as a large-distance expansion. The first terms in this expansion are:
\begin{align}
\label{eq:action_physical_gauge}
S = \sigma_0 R N_t + \frac{\sigma_0}{2} \int \mathrm{d}^2 \xi \left[\partial_\alpha X_i \partial_\alpha X^i + \frac{1}{8} ( \partial_\alpha X_i \partial_\alpha X^i )^2 -\frac{1}{4}( \partial_\alpha X_i \partial_\beta X^i )^2 + \dots \right]. 
\end{align}
It is important to note that, since the physical gauge discussed above is anomalous in $D\neq26$, in the three-dimensional case the action in eq.~\eqref{eq:action_physical_gauge} is only an effective description of the original Nambu--Got{\=o} action. One of the goals of this paper is to identify the leading corrections with respect to the physical gauge limit appearing at high orders in the low-energy expansion.

Moreover, it can be shown that the additional terms in the expansion of eq.~\eqref{eq:action_physical_gauge} beyond the Gau{\ss}ian one combine themselves so as to give an exactly integrable, irrelevant perturbation of the Gau{\ss}ian term, driven by a $\TTbar$ deformation of the quantum field theory of $(D-2)$ free bosons in two spacetime dimensions~\cite{Caselle:2013dra, Dubovsky:2017cnj, Dubovsky:2018bmo, Chen:2018keo}. Such deformations, built from the composite field obtained from the components of the energy-momentum tensor, were first discussed in detail in ref.~\cite{Zamolodchikov:2004ce}, and have since attracted considerable attention, since they can be studied analytically in terms of an inviscid Burgers equation~\cite{Zamolodchikov:2004ce, Smirnov:2016lqw, Cavaglia:2016oda, Conti:2018jho, Conti:2018tca, Conti:2019dxg} and have very interesting geometric implications~\cite{McGough:2016lol, Giveon:2017nie, Dubovsky:2017cnj, Dubovsky:2018bmo, Datta:2018thy, Aharony:2018bad, Kraus:2018xrn, Hartman:2018tkw, Bonelli:2018kik, Gorbenko:2018oov, Conti:2018tca, Dong:2018cuv, Cardy:2018sdv, Cardy:2019qao}. Thanks to the ``solvable'' nature of the $\TTbar$ deformation, the partition function of the model can be written explicitly and thus the correlator between two static color sources, that is the object of interest in this work, can be expressed in terms of a series of modified Bessel functions of the second kind. In $D$ spacetime dimensions the expression is the following
\begin{align}
\label{eq:polyakov_bessel}
G(R)=\sum_{n=0}^{\infty} w_n \frac{2R \sigma_0 L_t}{E_n}\left(\frac{\pi}{\sigma_0}\right)^{\frac{D-2}{2}} \left( \frac{E_n}{2\pi R} \right)^{\frac{D-1}{2}} K_{\frac{D-3}{2}}(E_nR),
\end{align}
which is consistent with earlier calculations based on the open-closed string duality~\cite{Luscher:2004ib} or on covariant quantization in the D$0$-brane formalism~\cite{Billo:2005iv}. Here $K_\rho(z)$ is the modified Bessel function of the second kind of order $\rho$ and argument $z$, $R$ denotes the distance between the static color sources, $L_t$ the extent of the Euclidean-time direction, and $w_n$ is the multiplicity of the closed string state that propagates from one Polyakov loop to the other. Note that the generating function for the latter is the Dedekind function describing the large-$R$ limit of eq.~\eqref{eq:polyakov_bessel}:
\begin{align}
\left( \prod_{r=1}^\infty \frac{1}{1-q^r} \right)^{D-2} = \sum_{k=0}^\infty w_k q^k.
\end{align}
Finally, the energy levels $E_n$ are given by
\begin{align}
E_n = \sigma_0 L_t \sqrt{1 + \frac{8 \pi}{\sigma_0 L_t^2} \left( n- \frac{D-2}{24} \right) }.
\end{align}
It is important to stress that eq.~\eqref{eq:polyakov_bessel} is only a large-distance expansion, and is valid only for separations between the sources larger than a critical radius $R_c$. In the framework of the Nambu-Got{\=o} action, the critical radius can be evaluated as $R_c=\sqrt{\frac{\pi(D-2)}{12\sigma_0}}$.

At large distances, the $G(R)$ correlator is dominated by the lowest state ($n=0$) and can be approximated as
\begin{align}
\label{eq:correlator_NG}
G(R) \simeq \left(\frac{1}{R} \right)^{\frac{D-3}{2}} K_{\frac{D-3}{2}} (E_0 R),
\end{align}
where the lowest energy state is given by
\begin{align}
\label{eq:ground_NG}
E_0 = \sigma_0 L_t \sqrt{1 - \frac{\pi (D-2)}{3 \sigma_0 L_t^2}}.
\end{align}
$E_0$ can be interpreted as the inverse of the correlation length $\xi_l$.

As we are interested in applying the EST predictions to the non-zero temperature regime of gauge theories, we define the system in a Euclidean space with a compact direction whose size, denoted as $L_t$, is the inverse of the temperature $T$. Then, we introduce the temperature-dependent string tension $\sigma(T)$, defined as 
\begin{align}
\label{eq:sigmaT}
\sigma(T) \equiv \frac{E_0}{L_t} = \sigma_0 \sqrt{1 - \frac{\pi(D-2)}{3 \sigma_0 L_t^2}}.
\end{align}
In this picture it is natural to define the temperature at which $\sigma(T)$ is vanishing as the critical temperature $T_{c,\mathrm{NG}}$~\cite{Pisarski:1982cn, Olesen:1985ej}:
\begin{align}
\label{eq:TCNG}
\frac{T_{c,\mathrm{NG}}}{\sqrt{\sigma_0}} = \sqrt{\frac{3}{\pi(D-2)}},
\end{align}
from which one can predict the critical exponent $\nu = 1/2$. 

However, this description for the deconfinement transition is not expected to be correct, as the critical index should instead be that of the symmetry-breaking phase transition of the $(D-1)$ dimensional spin model with symmetry group the center of the original gauge group (we will discuss this mapping in detail in section~\ref{sec:svetitsky_yaffe}). For instance, the deconfinement phase transition of the $\SU(2)$ lattice gauge theory in three dimensions, which is continuous, belongs to the same universality class of the symmetry-breaking phase transition of the two-dimensional Ising model, from which one has $\nu=1$.

Furthermore, while eq.~\eqref{eq:TCNG} is, in its simplicity, a surprisingly good approximation of the actual deconfinement temperatures of the gauge theories we intend to study, it fails to provide a quantitatively robust prediction of $T_c$, as we will analyze later in detail. These observations indicate that to obtain the correct EST describing the gauge theory, it is essential to go beyond the Nambu--Got{\=o} approximation.

\subsection{Beyond the Nambu--Got{\=o} approximation}
\label{sec:BNG}

Finding the correct terms of the EST action beyond the Nambu--Got{\=o} approximation is one of the major open challenges in the string description of confining gauge theories and a main goal of the present work.

From an effective-action perspective, one can start from the most general form of the action:
\begin{align}
 S = \sigma_0 R N_t + \frac{\sigma_0}{2} \int \mathrm{d}^2 \xi \left[\partial_\alpha X_i \partial_\alpha X^i + c_2 ( \partial_\alpha X_i \partial_\alpha X^i )^2 +c_3( \partial_\alpha X_i \partial_\beta X^i )^2 + \dots \right],
\end{align}
and then fix the coefficients order by order, e.g., from the results of lattice calculations. However, the ``low-energy universality'' of the EST~\cite{Luscher:2004ib, Meyer:2006qx, Aharony:2009gg, Aharony:2011gb, Gliozzi:2011hj, Gliozzi:2012cx, Dubovsky:2012sh, Aharony:2013ipa} implies that the $c_i$ coefficients cannot be arbitrary: they must satisfy a set of constraints in order to respect the Poincar{\'e} invariance of the gauge theory in the $D$-dimensional target space. In particular, it can be shown that the first correction with respect to the Nambu--Got{\=o} action, in the high-temperature regime which we are studying here, appears at order $1/L_t^{7}$ in the expansion of $E_0$ around the limit of an infinitely long string.

Moreover, studying the $2 \to 2$ scattering amplitude of the string excitations, in ref.~\cite{EliasMiro:2019kyf} it was found that the first two correction terms beyond the Nambu--Got{\=o} approximation (the $1/L_t^7$ and $1/L_t^9$ terms) depend on the same parameter, while the next independent parameter only appears at the $1/L_t^{11}$ order. Using notation similar to the one used in ref.~\cite{Baffigo:2023rin}, which in turn was inspired by the works based on the S-matrix approach~\cite{EliasMiro:2019kyf, EliasMiro:2021nul}, the expression for the non-universal corrections up to the order $1/L_t^{11}$ can be written as
\begin{align} 
\label{eq:ground_state_smatrix}
 E_0 \left(L_t\right) = \sigma_0 L_t \sqrt{1-\frac{\pi}{3 \sigma_0 L_t^2}} - \frac{32 \pi^6 \gamma_3}{225 \sigma_0^3 L_t^7} - \frac{64 \pi^7 \gamma_3}{675 \sigma_0^4 L_t^9}-\frac{\frac{2 \pi^8 \gamma_3}{45} + \frac{32768 \pi^{10} \gamma_5}{3969}}{\sigma_0^5 L_t^{11}}.
\end{align}
Following ref.~\cite{EliasMiro:2019kyf}, it is possible to set bounds on the values of these parameters from the bootstrap analysis; defining
\begin{align}
 \tilde{\gamma}_n=\gamma_n+(-1)^{(n+1) / 2} \frac{1}{n 2^{3 n-1}}
\end{align}
one finds that
\begin{align}
\label{eq:bound_bootstrap}
\tilde{\gamma}_3 \geq 0, \;\;\;\;\; \tilde{\gamma}_5 \geq 4 \tilde{\gamma}_3^2-\frac{1}{64} \tilde{\gamma}_3.
\end{align}
In particular, the most relevant constraint for the scope of this work is $\gamma_3 > -\frac{1}{768}$; note that the bound on $\gamma_5$ depends on the value of $\gamma_3$.

There is another term that contributes beyond the Nambu--Got{\=o} approximation, the so-called ``boundary term''. It can be shown that this term in the low-temperature regime is proportional to $1/R^4$, and hence is the dominant contribution. Its presence makes the detection of the corrections to the Nambu--Got{\=o} approximation at zero temperature very challenging, as they appear at order $1/R^7$, and, thus, are strongly suppressed and get masked by the boundary term. However, it can be shown that in the high-temperature regime and in the limit of very large separation of the two Polyakov loops, $R \gg L_t$, the boundary corrections become subleading, making it possible to access the more interesting bulk corrections~\cite{Caselle:2021eir}: this motivates our strategy to investigate the corrections to the Nambu--Got{\=o} action at temperatures close to the deconfinement phase transition.

\section{Lattice gauge theory setup}
\label{sec:lattice}

In this section we describe the setup of our numerical lattice simulations. We consider $\SU(N)$ Yang--Mills theories in $2+1$ spacetime dimensions, focusing on the cases of $N=3$ and $N=6$ color charges. Such theories are regularized on a lattice of $N_s^2 \times N_t$ cubic cells with a lattice spacing $a$ and periodic boundary conditions in the three main directions. The physical temperature $T$ is identified with the inverse of the extent of the system in the Euclidean-time direction, $T = 1/L_t = 1/(a N_t)$, while the extent of the system in the two remaining directions is denoted as $L_s = aN_s$. As we are interested in studying the system at relatively high temperatures, we chose $N_t \sim O(10) \ll N_s \sim O(100)$.

We used the discretization of the purely gluonic action due to Wilson~\cite{Wilson:1974sk}:
\begin{align}
S_W [U] = \beta \sum_{x, \mu < \nu} \left( 1 - \frac{1}{N} \real \Tr \Pi_{\mu \nu} (x) \right) ,
\end{align}
where the bare coupling $g$ appears in the Wilson parameter $\beta$, defined as $\beta=2N/(ag^2)$, and $\Pi_{\mu\nu}(x) = U_\mu(x) U_\nu(x+\hat{\mu}) U_\mu^\dagger(x+\hat{\nu}) U_\nu^\dagger(x)$ is the product of the $\SU(N)$ group elements (in the fundamental representation) along the edges of the $a \times a$ square in the $(\mu,\nu)$ plane, starting from the lattice site $x$.

It is very well known that these theories feature a phase transition associated with the spontaneous breaking of the $\Z_N$ center symmetry at a finite temperature $T_c$, that depends on $N$. In the $T < T_c$ region the theory is linearly confining, the Polyakov loop, defined as
\begin{align}
P(\Vec{x}) = \frac{1}{N} \Tr \left[ \prod_{t = 0}^{N_t} U_0\left(\Vec{x}, t \right) \right]
\end{align}
(where $\Vec{x}$ indicates the spatial coordinates of the loop and $U_0(\Vec{x}, t)$ a link variable in the Euclidean-time direction) has a vanishing expectation value in the thermodynamic limit, and the center symmetry is realized. On the other hand, in the $T > T_c$ regime the theory is in the deconfined phase, where center symmetry is spontaneously broken and the Polyakov loop has a non-zero expectation value. 

The temperature at which the phase transition takes place has been investigated for a broad range of lattice parameters; in particular, a precise determination of $\beta_c$ at different values of $N_t$ based on results from ref.~\cite{Liddle:2008kk} is given by the following formula~\cite{Caselle:2011fy}:
\begin{align}
\label{eq:scale_setting}
\frac{T}{T_c} = \frac{1}{N_t} \frac{\beta - 0.22N^2 + 0.5}{0.375N^2+0.13-0.211/N^2}.
\end{align}
We investigated the behavior of the system at temperatures in the proximity of the transition, but below $T_c$, roughly in the $0.5 < T/T_c < 1.0$ region.

In this temperature range we studied the two-point correlation function of Polyakov loops, as a function of their distance $R$:
\begin{align}
G(R) = \frac{1}{2N_s^2} \left\langle \sum_{\Vec{x}, \hat{k}} P(\Vec{x}) P(\Vec{x} + \, \hat{k} R ) \right\rangle.
\end{align}
Here the $\langle \cdots \rangle$ notation indicates the mean over the sampled lattice configurations; the correlator is averaged also on the spatial volume, to increase the statistical precision of the results. The physical relevance of $G(R)$ is due to its relation to the potential $V(R, N_t)$ between probe sources in the fundamental representation of the gauge group:
\begin{align}
\label{eq:potential_defined}
V(R, L_t) \equiv -\frac{1}{L_t} \ln G(R).
\end{align}

\subsection{Detection of BNG corrections from lattice simulations} 

In the confining phase and for sufficiently large spatial separations, the potential grows linearly with $R$; this is perfectly encapsulated by the long-distance behavior of the EST prediction of eq.~\eqref{eq:correlator_NG}, which simply reads
\begin{align}
G(R) \propto \exp\left[- \sigma(T) L_t R\right],
\end{align}
where $\sigma(T)$ is the finite-temperature string tension defined in eq.~\eqref{eq:sigmaT}.

We are now interested in the fine details of the behavior of the correlator $G(R)$ and we follow the same approach of the study on the $\SU(2)$ gauge theory in $2+1$ dimensions that was presented in ref.~\cite{Caristo:2021tbk}: we estimate the EST correction beyond the Nambu--Got{\=o} approximation by studying the behavior of the ground-state energy $E_0$ when the deconfinement transition temperature is approached from below. At a fixed value of $\beta$, the temperature is changed by varying $N_t$: we performed numerical simulations at different values of $\beta$ and we determined the ground state energy $E_0$ as the inverse of the longest correlation length in the system, denoted as $\xi_l$. In our lattice simulations we computed the values of $G(R)$ for $R \le L_s/2$.

Starting from the EST prediction in eq.~\eqref{eq:correlator_NG}, we assume the lattice results for the Polyakov loop correlator to be described by the functional form
\begin{align}
\label{eq:high_dist_corr}
G(R) = k_l \, \left[K_0\left(\frac{R}{\xi_l}\right) + K_0\left(\frac{L_s - R}{\xi_l}\right)\right].
\end{align}
The second term on the right-hand-side of eq.~\eqref{eq:high_dist_corr} takes into account the leading effect of the periodic copies of the system, due to the boundary conditions: the inclusion of this term is necessary, in order to treat properly cases in which the correlation length has an extent almost comparable with the linear size of the lattice, as can occur at temperatures close to the deconfinement transition. As we will discuss later in more detail, we repeated our simulation on lattices of larger spatial volumes for some $N_t$ close to the critical point, to make sure that effects due to the finiteness of the spatial extent of the system are properly accounted for in our analysis. Equation~\eqref{eq:high_dist_corr} can be considered as a sufficiently accurate approximation at least for $R > \xi_l$, which is a region where the contribution of higher-energy states $E_n$, for $n>0$, can be neglected. We carefully tested that this is the case for all of the lattice results that we discuss in the following. Moreover, as we already remarked, the EST description is valid only for distances larger than a critical radius $R_c$; the constraint $R_c<\xi_l$ is fulfilled for all of our data.

We extract the ground energy $E_0$ from the inverse of the correlation length $\xi_l$ and we study its behavior in $N_t$. From the considerations of section~\ref{sec:BNG}, we know that the first correction for $E_0$ to the Nambu--Got{\=o} approximation is predicted to arise at the $1/N_t^7$ order, while the second and the third at order $1/N_t^9$ and $1/N_t^{11}$, according to eq.~\eqref{eq:ground_state_smatrix}. Thus, we assume the following form for the $N_t$ dependence of the ground state $E_0$:
\begin{align}
\label{eq:BNG_correction}
aE_0(N_t) = N_t\sigma_0 a^2 \sqrt{1-\frac{\pi}{3N_t^2 \; \sigma_0 a^2}} + \frac{k_4}{(\sigma_0 a^2)^3 N_t^7} + \frac{2 \pi k_4}{3(\sigma_0 a^2)^4 N_t^9} + \frac{5 \pi^2 k_4}{16 (\sigma_0 a^2)^5 N_t^{11}} + \frac{k_5}{(\sigma_0 a^2)^5 N_t^{11}}.
\end{align}
Moreover, since there may also be higher-order corrections, it is natural to truncate for consistency the Nambu--Got{\=o} prediction to the corresponding order of the correction, obtaining the following parametrization:
\begin{align}
\label{eq:BNG_groundstate_Nt11}
aE_0(N_t) = \mathrm{Taylor}_6 (E_0) + \frac{k_4}{(\sigma_0 a^2)^3 N_t^7} + \frac{2 \pi k_4}{3(\sigma_0 a^2)^4 N_t^9} + \frac{5 \pi^2 k_4}{16 (\sigma_0 a^2)^5 N_t^{11}} + \frac{k_5}{(\sigma_0 a^2)^5 N_t^{11}} ,
\end{align}
where
\begin{align}
\begin{aligned}
\label{eq:NG_taylor}
\mathrm{Taylor}_6(E_0) \equiv N_t\sigma_0 a^2 - \frac{\pi}{6N_t} - \frac{\pi^2}{72 \sigma_0 a^2 \, N_t^3} - \frac{\pi^3}{432 (\sigma_0 a^2)^2 N_t^5} - \frac{5\pi^4}{10361 8(\sigma_0 a^2)^3 N_t^7} \\ 
- \frac{7\pi^5}{62208(\sigma_0 a^2)^4 N_t^9} - \frac{21 \pi^6}{746496 (\sigma_0 a^2)^5 N_t^{11}} .
\end{aligned}
\end{align}
However, as we will discuss in the following sections, the values of the parameters $k_4$ and $k_5$ are strongly affected by the truncation of both the order of the correction and the order at which we truncate the Nambu--Got{\=o} prediction. For this reason, to obtain a reliable estimate of these parameters it is necessary to carry out a careful analysis of systematic uncertainties. Note that in the previous work on the $\SU(2)$ gauge theory~\cite{Caristo:2021tbk}, such analysis was not performed, and the expression for the corrections beyond the Nambu--Got{\=o} action were known only up to $1/N_t^7$ order. For this reason, in this work we also carry out an improved determination of $k_4$ and an estimate of $k_5$ for the $\SU(2)$ gauge theory.

\section{The Svetitsky--Yaffe mapping}
\label{sec:svetitsky_yaffe}

For Yang--Mills theories undergoing a \emph{continuous} thermal deconfining phase transition, close to the critical temperature it is possible to describe the physics on distance scales larger than the correlation length in terms of a low-energy effective theory, whereby the system becomes independent of the microscopic details of the underlying gauge theory. This is the basic idea underlying a famous conjecture that was put forward long ago by Svetitsky and Yaffe~\cite{Svetitsky:1982gs}. According to this conjecture, the degrees of freedom associated to the Polyakov loops in a Yang--Mills theory in $(d+1)$ dimensions can be described in terms of a spin model, characterized by a global symmetry with respect to the center of the gauge group of the original Yang--Mills theory, and defined in $d$ spatial dimensions.

This gauge-spin mapping has several interesting features: first, the deconfined (high-temperature) phase of the original gauge theory corresponds to the ordered (low-temperature) phase of the spin model. This correspondence between the high- and low-temperature phases stems from the fact that both phases correspond to the broken-symmetry regime. A second important consequence of this correspondence consists in the fact that the correlator of Polyakov loops in the confining phase is expected to be described in terms of the spin-spin correlation function in the disordered phase of the spin model. Finally, the plaquette operator in the gauge theory (which encodes the local Euclidean-action density) is mapped into the energy operator of the effective spin model. 

In ref.~\cite{Caristo:2021tbk} this conjecture was already tested for the $\SU(2)$ gauge theory in $2+1$ dimensions; according to the mapping described above, the predicted universality class is the one of the two-dimensional Ising model, which is an exactly solved model~\cite{Onsager:1943jn, Kaufman:1949ks, Kac:1952tp, Hurst:1960oae, Schultz:1964fv}, and high-precision numerical simulations of the $\SU(2)$ gauge theory confirmed this conjecture.

In this work, similarly, we will test the conjectured correspondence between the $\SU(3)$ Yang--Mills theory in three dimensions (which, like the $\SU(2)$ theory, also has a continuous thermal deconfinement phase transition) and the three-state Potts model in two dimensions. Note that the Svetitsky--Yaffe conjecture does not apply to the $\SU(6)$ gauge theory in three dimensions, since in this case the transition is of the first order~\cite{Liddle:2008kk}. For the $\SU(3)$ gauge theory, we will study the Svetitsky--Yaffe mapping by comparing the behavior of the Polyakov loop correlator with the spin-spin correlator of the Potts model, which was analyzed in ref.~\cite{Caselle:2005sf}.

An important motivation to consider the Svetitsky--Yaffe conjecture in our present study is that it allows us to obtain a numerical estimate of the correlation length of the system in an independent way, using only the values of the Polyakov-loop correlators at short distances. An agreement between this short-distance estimate of the correlation length and the one deduced from the EST model represents a robust consistency test of our results.

\subsection{Two-point correlation function in the three-state Potts model} 

As mentioned above, the Svetitsky--Yaffe conjecture predicts that, close to the deconfinement temperature, the confining phase of $\SU(3)$ Yang--Mills theory in $2+1$ dimensions is mapped to the disordered phase of the three-state Potts model in two dimensions. As we are close to criticality, we can rely on the continuum description in terms of the thermal perturbation of the conformal field theory (CFT) describing the critical behavior of the three-state Potts model, which is a minimal model with central charge $c=4/5$,
\begin{align} 
\label{eq:Stau} 
S = S_{\mathrm{CFT}} + \tau \int \mathrm{d}^2 x \; \epsilon(x) 
\end{align} 
and belongs to the class of integrable quantum field theories~\cite{Zamolodchikov:1989zs}. The Svetitsky--Yaffe conjecture also maps the Polyakov-loop correlator of the gauge theory to the spin-spin correlation function of the Potts model, denoted as $\langle \sigma \overline{\sigma} \rangle$; to have good analytical control of the latter, one can use the framework of conformal perturbation theory~\cite{Zamolodchikov:1990bk, Guida:1995kc, Guida:1996ux, Guida:1996nm}, which are expected to hold for distances much shorter than the largest correlation length of the system.
 
Let us review the main results of conformal perturbation theory for the $\langle \sigma(x) {\overline{\sigma}}(0) \rangle$ correlation function. Following the notation used in the literature on the subject, this correlator can be written as
\begin{align} 
G_\sigma(x) = \langle \sigma(x) {\overline{\sigma}}(0) \rangle = 
\sum_p {\mathcal C}_{\sigma {\overline{\sigma}}}^{[\phi_p]} (x; \tau) 
\langle [\phi_p] \rangle ,
\end{align} 
where the sum over $p$ ranges over all conformal families allowed by the operator product expansion of $\sigma \overline{\sigma}$. The Wilson coefficients ${\mathcal C}_{\sigma {\overline{\sigma}}}^{[\phi_p]}$ can be calculated perturbatively in the coupling constant $\tau$. Their Taylor expansion in powers of $\tau$ is
\begin{align} 
 {\mathcal C}_{\sigma {\overline{\sigma}}}^{[\phi_p]}(x;\tau) = \sum_k 
 \frac{\tau^k}{k!} \partial^k_\tau 
 {\mathcal C}_{\sigma {\overline{\sigma}}}^{[\phi_p]}(x;0). 
\end{align} 
It is possible to show that the derivatives of the Wilson coefficients appearing in the previous expression can be written in terms of multiple integrals of the conformal correlators 
\begin{align} 
\partial^k_\tau 
 {\mathcal C}_{\sigma {\overline{\sigma}}}^{[\phi_p]}(x;0) = 
(-1)^k \int^\prime \mathrm{d}^2 z_1 \dots \mathrm{d}^2 z_k \, 
 \langle \sigma(x) {\overline{\sigma}(0)} \epsilon(z_1) \dots \epsilon(z_k) [\phi_p](\infty) 
 \rangle_{\mathrm{CFT}} . 
\end{align} 
For all the details we address the interested reader to ref.~\cite{Guida:1995kc}. 
 
Another important ingredient entering the perturbative expansion of the correlator is represented by the vacuum expectation values $\langle [\phi_p] \rangle$; they are of non-perturbative nature and were computed in refs.~\cite{Lukyanov:1996jj, Fateev:1997yg, Fateev:1998xb} for a wide class of theories, including various integrable perturbations of the minimal models. 
 
The leading term of the perturbative expansion is given by the two-point conformal correlator, which corresponds to the choice $k=0$ and $\phi_p= \ide$:
\begin{align} 
\label{eq:2pt} 
{\mathcal C}_{\sigma {\overline{\sigma}}}^{ \ide}(x;0)= \frac{1}{|x|^{4/15}},
\end{align} 
where the so-called ``conformal normalization'', $C_{\sigma {\overline{\sigma}}}^{\ide}(x;0)|_{x=1}=1$, is assumed. The first few subleading terms are 
\begin{align} 
G_\sigma(x) = \langle \sigma(x) {\overline{\sigma}}(0) \rangle = 
{\mathcal C}_{\sigma {\overline{\sigma}}}^{ \ide}(x;\tau) + 
{\mathcal C}_{\sigma {\overline{\sigma}}}^{ \epsilon}(x;\tau) 
\langle \epsilon \rangle+ \dots ,
\end{align}
where
\begin{align}
{\mathcal C}_{\sigma {\overline{\sigma}}}^{ \ide}(x;\tau) & = {\mathcal C}_{\sigma {\overline{\sigma}}}^{ \ide}(x;0) + \tau \, \partial_\tau {\mathcal C}_{\sigma {\overline{\sigma}}}^{ \ide}(x;0) + \dots \nonumber \\ 
{\mathcal C}_{\sigma {\overline{\sigma}}}^{ \epsilon}(x;\tau) & = 
{\mathcal C}_{\sigma {\overline{\sigma}}}^{ \epsilon }(x;0) + \dots
\end{align}
give corrections up to $\tau$. The explicit expression of the various contributions is, together with eq.~\eqref{eq:2pt},
\begin{align}
\partial_\tau {\mathcal C}_{\sigma {\overline{\sigma}}}^{ \ide}(x;0) & = - \int \mathrm{d}^2 z \langle \sigma(x) {\overline{\sigma}(0)} \epsilon(z) \rangle_{\mathrm{CFT}} = -C_{\sigma {\overline{\sigma}}}^\epsilon \, |x|^{14/15} \int \mathrm{d}^2 y |y|^{-4/5} |1-y|^{-4/5} \nonumber \\ 
&= \sin \left(\frac{4\pi}{5}\right) \left| \frac{\Gamma(-1/5)\Gamma(3/5)}{\Gamma(2/5)} 
\right|^2 C_{\sigma {\overline{\sigma}}}^\epsilon \, |x|^{14/15} 
\nonumber \\ 
{\mathcal C}_{\sigma {\overline{\sigma}}}^{ \epsilon }(x;0) &= C_{\sigma {\overline{\sigma}}}^\epsilon \,|x|^{8/15} ,
\end{align} 
where the Wilson coefficient 
\begin{align} 
C_{\sigma {\overline{\sigma}}}^\epsilon = \sqrt{\frac{\cos\left(\frac{\pi}{5}\right)}{2}}\, 
 \frac{\Gamma^2(3/5)}{\Gamma(2/5)\Gamma(4/5)} = 0.546178 \dots 
\end{align} 
can be found combining the results of ref.~\cite{Dotsenko:1985hi} with those from ref.~\cite{Klassen:1991dz} (see also ref.~\cite{McCabe:1995uq}). The integral appearing in 
$\partial_\tau {\mathcal C}_{\sigma {\overline{\sigma}}}^{ \ide}(x;0) $ is well known: it is a particular case of
\begin{align}
{\mathcal Y}_{a,b} = \int \mathrm{d}^2 z |z|^{2a} |1-z|^{2b} = \frac{\sin \left( \pi (a+b)\right) \, \sin (\pi b)}{\sin (\pi a)} 
 \left| \frac{\Gamma(-a-b-1)\Gamma(b+1)}{\Gamma(-a)} 
\right|^2 
\end{align} 
and its numerical value is ${\mathcal Y}_{-\frac25,-\frac25} = -8.97743 \dots$. 
 
Finally, let us discuss the other non-perturbative quantities required for our calculation, namely, the vacuum expectation value of the perturbing operator $\epsilon(x)$, and the relation between the coupling constant and the mass of the fundamental particle. The latter is given by~\cite{Fateev:1993av}:
\begin{align} 
\label{kappa}
\tau = \kappa\, m^{6/5}, \ \ \ \ \kappa = 0.164303 \dots ,
\end{align} 
whereas the former can be easily computed starting from the knowledge of the vacuum energy density~\cite{Zamolodchikov:1989cf}:
\begin{align} 
\varepsilon_0 = - \frac{\sqrt 3}{6} m^2, 
\end{align} 
which is related to the vacuum expectation value of the perturbing operator through 
\begin{align}
\label{eq:epsilon_vev}
\langle \epsilon \rangle = \partial_\tau \varepsilon_0 = 
A_\epsilon\, \tau^{2/3} . 
\end{align}
In eq.~\eqref{eq:epsilon_vev} the amplitude $A_\epsilon$ of the energy operator appears; it is a non-universal parameter, which depends on the specific microscopic realization of the underlying model, and thus must be evaluated using Monte Carlo simulations or other approaches. For the two-dimensional three-state Potts model, this parameter was computed in ref.~\cite{Caselle:2005sf} with the result $A_\epsilon =- 9.761465\dots$, leading to
\begin{align} 
\label{eq:epsilon_vev_bis}   
\langle \epsilon \rangle = - 9.761465\dots \tau^{2/3} =- 2.92827\dots m^{4/5}.   
\end{align}   

The perturbative series can then be recast in the following form: 
\begin{align}
G_\sigma(x) & = \frac{1}{|x|^{4/15}} \left(1 + C_{\sigma {\overline{\sigma}}}^\epsilon A_\epsilon \, \tau^{2/3} |x|^{4/5} - {\mathcal Y}_{-\frac25,-\frac25} C_{\sigma {\overline{\sigma}}}^\epsilon \, \tau |x|^{6/5} +\dots \right) \nonumber \\ 
& = \frac{1}{|x|^{4/15}} \left(1 + C_{\sigma {\overline{\sigma}}}^\epsilon A_\epsilon \, u^{2/3} - {\mathcal Y}_{-\frac25,-\frac25} C_{\sigma {\overline{\sigma}}}^\epsilon \, u 
+\dots \right) \nonumber \\
& = \frac{1}{|x|^{4/15}} \left(1 + C_{\sigma {\overline{\sigma}}}^\epsilon A_\epsilon \kappa^{2/3} \, r^{4/5} - {\mathcal Y}_{-\frac25,-\frac25} C_{\sigma {\overline{\sigma}}}^\epsilon \, \kappa \, r^{6/5} +\dots \right), \nonumber 
\end{align} 
having set $u=\tau |x|^{6/5}$, and $r = m |x|$. This expression can be rewritten in terms of the dimensionless variable $r$ as
\begin{align}
\label{eq:spinspincorr}
{\tilde G_\sigma (r)} = m^{-4/15} \langle \sigma(x) {\overline{\sigma}}(0) \rangle = \frac{1}{r^{4/15}} \left( 1+g_1 r^{4/5}+g_2 r^{6/5}+ \dots \right),
\end{align} 
where  
\begin{align}
g_1=C_{\sigma {\bar \sigma}}^\epsilon  A_\epsilon  \kappa^{2/3}, \ \ \ \ g_2=-{\mathcal Y}_{-\frac25,-\frac25} C_{\sigma {\bar \sigma}}^\epsilon    
\kappa =0.805622 \dots .
\end{align}

The last ingredient to construct the $\langle \sigma(x) {\overline{\sigma}}(0) \rangle$ correlator corresponding to the Polyakov-loop correlator of the $\SU(3)$ gauge theory is the amplitude $A_\sigma$ associated with the spin operator. While the normalization used in eq.~\eqref{eq:2pt} would correspond to $A_\sigma=1$, this parameter is non-universal, hence it will be extracted from our lattice data. Thus, the expression that we use to fit our lattice data at short distances reads
\begin{align} 
\label{eq:low_dist_corr_1}
G(R) = \frac{A_\sigma^2}{|r|^{4/15}} \left( 1 + g_1 r^{4/5} + g_2 r^{6/5} \right),
\end{align}
where $r=R/\xi_l$, and we fix $g_2 = 0.805622 \dots$, leaving the $A_\sigma$ amplitude, the correlation length $\xi_l$, and the $g_1$ constant as the free parameters of the fits. In particular, it will be interesting to compare the fit results for $g_1$ with the value this quantity has in the Potts model, which can be evaluated exactly and is
\begin{align}   
g_1=C_{\sigma {\bar \sigma}}^\epsilon  A_\epsilon  \kappa^{2/3}=-1.59936\dots . 
\end{align}

\section{Numerical results}
\label{sec:results}

In this section we present the results of our Monte~Carlo simulations of $\SU(3)$ and $\SU(6)$ lattice gauge theories. The simulations were performed using either the code employed in ref.~\cite{Bonati:2021vbc}, or the code originally developed for the studies presented in refs.~\cite{Panero:2009tv,Mykkanen:2012ri}. The two simulation codes are in perfect agreement on standard benchmark tests and the results of our simulations are totally consistent.

\subsection{$\SU(3)$ Yang--Mills theory: extracting the ground state energy $E_0$ from long-distance fits}
\label{sec:su3res}

We performed several simulations of $\SU(3)$ pure gauge theory at different values of $\beta$, $N_t$ and $N_s$. Using eq.~\eqref{eq:scale_setting}, we tuned the values of $\beta$ to fix the lattice spacing to four different values, namely $a=1/(6.5 T_c)$, $a=1/(8.5 T_c)$, $a=1/(9.5 T_c)$ and $a=1/(10.5 T_c)$. Then, we varied the values of $N_t$ in order to move closer to or away from the transition. In simulations performed closer to the transition, $N_s$ was increased, too, in order to accommodate for the larger correlation lengths. The main technical details of these simulations are reported in table~\ref{tab:su3tableconf}.

\begin{table}[H]
 \centering
 \begin{subtable}[b]{0.45\textwidth}
 \centering
 \begin{tabular}{|l|l|l|l|l|}
 \hline
 $\beta$ & $N_t$ & $N_s$ & $T/T_c$ & $n_{\mathrm{conf}}$ \\ \hline
 \multirow{8}{*}{23.11} & 7 & 160 & 0.93 & $1.2\times10^5$ \\ \cline{2-5} 
 & 8 & 160 & 0.81 & $1.4\times10^5$ \\ \cline{2-5} 
 & 9 & 96 & 0.72 & $3.9\times10^5$ \\ \cline{2-5} 
 & 10 & 96 & 0.65 & $3.9\times10^5$ \\ \cline{2-5} 
 & 11 & 96 & 0.59 & $2.3\times10^5$ \\ \cline{2-5} 
 & 12 & 96 & 0.54 & $2.2\times10^5$ \\ \cline{2-5} 
 & 13 & 96 & 0.50 & $1.1\times10^5$ \\ \cline{2-5} 
 & 14 & 96 & 0.47 & $1.5\times10^5$ \\ \hline
 \end{tabular}
 \caption{$\beta=23.11$, $1 / (a \, T_c) = 6.5$.}
 \label{su3tableconf1}
 \end{subtable}
 \hfill
 \begin{subtable}[b]{0.45\textwidth}
 \centering
 \begin{tabular}{|l|l|l|l|l|}
 \hline
 $\beta$ & $N_t$ & $N_s$ & $T/T_c$ & $n_{\mathrm{conf}}$ \\ \hline
 \multirow{8}{*}{29.82} & 9 & 240 & 0.95 & $6\times10^4$ \\ \cline{2-5} 
 & 10 & 160 & 0.85 & $9.6\times10^4$ \\ \cline{2-5} 
 & 11 & 96 & 0.78 & $2.3\times10^5$ \\ \cline{2-5} 
 & 12 & 96 & 0.71 & $2.2\times10^5$ \\ \cline{2-5} 
 & 13 & 96 & 0.67 & $2.2\times10^5$ \\ \cline{2-5} 
 & 14 & 96 & 0.61 & $1.9\times10^5$ \\ \cline{2-5} 
 & 15 & 96 & 0.57 & $2.2\times10^5$ \\ \cline{2-5} 
 & 16 & 96 & 0.53 & $1.7\times10^5$ \\ \hline
 \end{tabular}
 \caption{$\beta=29.82$, $1 / (a \, T_c) = 8.5$.}
 \label{su3tableconf2}
 \end{subtable}

 \vspace{0.5cm}
 
 \begin{subtable}[b]{0.45\textwidth}
 \centering
 \begin{tabular}{|l|l|l|l|l|}
 \hline
 $\beta$ & $N_t$ & $N_s$ & $T/T_c$ & $n_{\mathrm{conf}}$ \\ \hline
 \multirow{8}{*}{33.18} & 10 & 240 & 0.95 & $9.6\times10^4$ \\ \cline{2-5} 
 & 11 & 160 & 0.87 & $8.6\times10^4$ \\ \cline{2-5} 
 & 12 & 96 & 0.80 & $2.5\times10^5$ \\ \cline{2-5} 
 & 13 & 96 & 0.73 & $2\times10^5$ \\ \cline{2-5} 
 & 14 & 96 & 0.68 & $2.3\times10^5$ \\ \cline{2-5} 
 & 15 & 96 & 0.64 & $2\times10^5$ \\ \cline{2-5} 
 & 16 & 96 & 0.60 & $1.8\times10^5$ \\ \cline{2-5} 
 & 17 & 96 & 0.56 & $1.7\times10^5$ \\ \hline
 \end{tabular}
 \caption{$\beta=33.18$, $1 / (a \, T_c) = 9.5$.}
 \label{su3tableconf3}
 \end{subtable}
 \hfill
 \begin{subtable}[b]{0.45\textwidth}
 \centering
 \begin{tabular}{|l|l|l|l|l|}
 \hline
 $\beta$ & $N_t$ & $N_s$ & $T/T_c$ & $n_{\mathrm{conf}}$ \\ \hline
 \multirow{8}{*}{36.33} & 11 & 240 & 0.95 & $6.2\times10^4$ \\ \cline{2-5} 
 & 12 & 160 & 0.87 & $1.1\times10^5$ \\ \cline{2-5} 
 & 13 & 96 & 0.81 & $3\times10^5$ \\ \cline{2-5} 
 & 14 & 96 & 0.75 & $3\times10^5$ \\ \cline{2-5} 
 & 15 & 96 & 0.70 & $3\times10^5$ \\ \cline{2-5} 
 & 16 & 96 & 0.66 & $2.7\times10^5$ \\ \cline{2-5} 
 & 17 & 96 & 0.62 & $2.6\times10^5$ \\ \cline{2-5} 
 & 18 & 96 & 0.58 & $2.2\times10^5$ \\ \hline
 \end{tabular}
 \caption{$\beta=36.33$, $1 / (a \, T_c) = 10.5$.}
 \label{su3tableconf4}
 \end{subtable}
 \caption{Parameters of our simulations of the $\SU(3)$ Yang--Mills theory, for different values of the lattice spacing $a$.}
 \label{tab:su3tableconf}
\end{table}

For each simulation, we performed a fit to the computed values of $G(R)$ using eq.~\eqref{eq:high_dist_corr}, considering only values with $R > \xi_l$, where we expect the subleading, exponentially decaying terms of the correlator to be negligible. The results of the long-distance best fits according to eq.~\eqref{eq:high_dist_corr} are reported in tables~\ref{tab:su3tableres1}, \ref{tab:su3tableres2}, \ref{tab:su3tableres3}, and~\ref{tab:su3tableres4} in appendix~\ref{sec:appendix}; furthermore, in fig.~\ref{fig:high_dist_fit_Nt7} the result of the fit of the correlator for $N_t=7$ at $\beta=23.11$ is shown. The estimate of $G(R)$ at different values of $R$ computed on the same configurations exhibits non-negligible cross-correlations, which are taken into account in the fit procedure and in the computation of the $\chi^2$. The values of the reduced $\chi^2$ that we obtained, which are of order unity, indicate the good quality of the fits.

\begin{figure}[h]
 \centering 
 \includegraphics[width=0.70\textwidth]{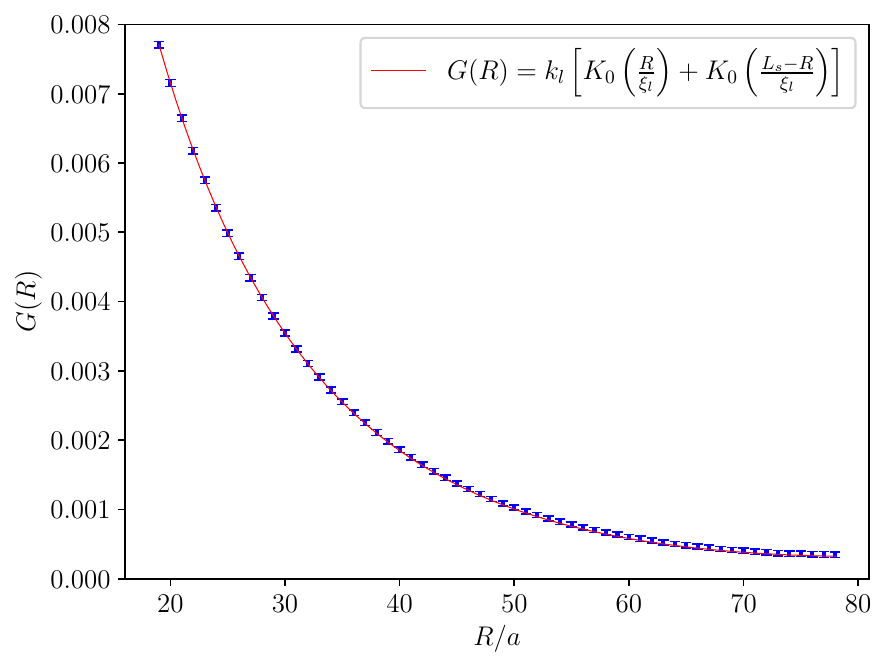}
 \caption{Best fits of the $\SU(3)$ data at $N_t=7$, $\beta=23.11$ according to eq.~\eqref{eq:high_dist_corr}.}
 \label{fig:high_dist_fit_Nt7}
\end{figure}

Next, we extract the value of the correlation length $\xi_l$ and we obtain the ground state energy as the inverse of the correlation length $E_0 = 1 / \xi_l$ for each value of $\beta$ and $N_t$. Since all simulations are statistically independent from each other, these values are not affected by cross-correlations, simplifying the analysis. The results of these fits are reported in table~\ref{tab:short_long}.

\subsection{$\SU(3)$ Yang--Mills theory: testing the Svetitsky--Yaffe conjecture and extracting the ground state energy $E_0$ from short-distance fits}
\label{sec:su3res3}

We now proceed to test the mapping between the thermal deconfinement transition of the $\SU(3)$ Yang--Mills theory in $2+1$ dimensions and the three-state Potts model in two dimensions, described in section~\ref{sec:svetitsky_yaffe}.

The Svetitsky--Yaffe mapping allows one to extract the ground-state energy from the short-distance behavior of the correlator (as was already done for the $\SU(2)$ gauge theory in ref.~\cite{Caristo:2021tbk}), using eq.~\eqref{eq:low_dist_corr_1}.

\begin{figure}[H]
    \centering   
    \includegraphics[width=0.70\textwidth]{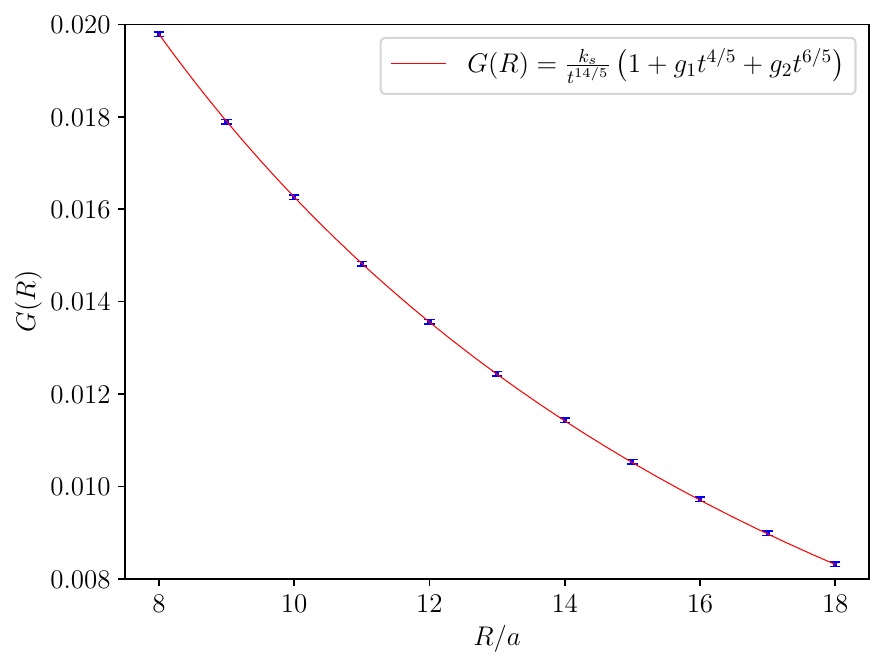}
    \caption{Best fits of the data at $N_t=7$, $\beta=23.11$ according to eq.~\eqref{eq:low_dist_corr_1}.}
    \label{fig:low_dist_fit_Nt7}
\end{figure}

We considered the results for $G(R)$ at the four $\beta$ values in table~\ref{tab:su3tableconf} at the temperatures with the largest correlation length, finding the results reported in table~\ref{tab:short_long}. In fig.~\ref{fig:low_dist_fit_Nt7} we also show the best fit of the data for $\beta=23.11$ and $N_t=7$ to eq.~\eqref{eq:low_dist_corr_1}.

The values of the correlation length $\xi_l$ obtained from the short-range fits are in good agreement (within less than three standard deviations) with those obtained by fitting the large distance data with the EST functional form of eq.~\eqref{eq:high_dist_corr} (which we also report in table~\ref{tab:short_long}, for comparison). This agreement confirms the robustness of our determination of the ground-state energy, and gives confidence on the procedure to determine BNG corrections.

Even though it is not a universal quantity, we expect the value of $g_1$ to be the same for each simulation, and not to depend on the lattice spacing or on the temperature. We expect this to hold up to small deviations due to further corrections to eq.~\eqref{eq:low_dist_corr_1}, which become important at lower temperatures (i.e., away from the phase transition). We tested this hypothesis by means of a combined fit, using a single $g_1$ parameter for all of the data above a threshold temperature, while leaving the values of $k_s$ and of $\xi_l$ as different free parameters for each simulation. This fit is found to yield an acceptable result, since for each data set the contribution to the $\chi^2$ of the combined fit is approximately equal to the number of fitted points minus two (i.e., to the number of free parameters of each fit). From the combined fit, we find the numerical value $g_1\approx-1.55$,\footnote{Including all data with $T / T_c \ge 80\%$ we obtain $g_1 = -1.55938(85)$, but the result appears to be quite sensitive to the datum at the lowest temperature included in the fit, suggesting the presence of a non-negligible systematic uncertainty.} which is similar to the value found for the Potts model in ref.~\cite{Caselle:2005sf}.

These results strongly support the validity of the Svetitsky--Yaffe mapping in the region of parameters that we study. It is also interesting to note how close the non-universal amplitude $A_\epsilon$ extracted from our data is to the one of the Potts model. While in principle there is no reason to expect $A_\epsilon$ to take the same value in the gauge theory and in the spin model, it is interesting to note that this similarity of the numerical values could, in principle, allow one to extract reliable quantitative information also for more complex correlation functions of the gauge theory, such as those involving more than two Polyakov loops, and/or products of different operators.

\subsection{Corrections beyond the Nambu--Got{\=o} string in $\SU(3)$ Yang--Mills theory}
\label{sec:su3res4}

First, we observe that the ground state values we obtained cannot be explained in terms of the temperature dependence predicted for the Nambu--Got{\=o} string: the data are not compatible with the functional form of eq.~\eqref{eq:ground_NG}. Fits performed under this assumption use $\sigma_0$ as the only free parameter and inevitably lead to $\chi^2$ of order $10$ times larger than the number of degrees of freedom. For this reason, we included in our fits the corrections parameterized by $k_4$ and we performed the best fits of $E_0$ for the four different lattice spacings separately: the results of this fit are reported in table~\ref{tab:BNG_SU3}.

\begin{table}[H]
\centering
\begin{tabular}{|l|l|l|l|l|l|l|}
\hline
 & $\beta$ & $N_{t,\mathrm{min}}$ & $N_{t,\mathrm{max}}$ & $k_4$ & $\sigma_0 a^2$ & $\chi^2/N_{\mathrm{d.o.f.}}$ \\ \hline
\multirow{ 4}{*}{up to $N_t^{-7}$} & 23.11 & 7 & 14 & -0.229(7) & 0.024713(14) & 1.3 \\ \cline{2-7}
& 29.82 & 10 & 16 & -0.218(16) & 0.014241(14) & 2.1 \\ \cline{2-7}
& 33.18 & 11 & 17 & -0.186(18) & 0.011305(13) & 1.3 \\ \cline{2-7}
& 36.33 & 12 & 18 & -0.22(2) & 0.009352(14) & 1.2 \\ \hline\hline
\multirow{ 4}{*}{up to $N_t^{-9}$} & 23.11 & 7 & 14 & -0.074(2) & 0.024665(13) & 3 \\ \cline{2-7} 
& 29.82 & 9 & 16 & -0.080(5) & 0.014225(12) & 2.6 \\ \cline{2-7} 
& 33.18 & 10 & 17 & -0.069(5) & 0.011295(12) & 1.2 \\ \cline{2-7} 
& 36.33 & 11 & 18 & -0.081(6) & 0.009337(12) & 1.4 \\ \hline
\end{tabular}\\

\vspace{0.5cm}

\begin{tabular}{|l|l|l|l|l|l|l|l||l|}
\hline
 &$\beta$ & $N_{t,\mathrm{min}}$ & $N_{t,\mathrm{max}}$ & $k_4$ & $k_5$ & $\sigma_0 a^2$ & $\chi^2/N_{\mathrm{d.o.f.}}$& $\sigma_0 a^2$ from ref.~\cite{Teper:1998te} \\ \hline
\multirow{ 4}{*}{up to $N_t^{-11}$}&23.11 & 7 & 14 & -0.126(17) & 0.63(12) & 0.024704(18) & 1.9 & 0.024701(80) \\ \cline{2-9}
&29.82 & 9 & 16 & -0.10(3) & 0.42(18) & 0.014233(16) & 2.9 & 0.014213(51) \\ \cline{2-9}
&33.18 & 10 & 17 & -0.04(3) & 0.1(2) & 0.011286(16) & 1.4 & 0.011339(53) \\ \cline{2-9}
&36.33 & 11 & 18 & -0.10(3) & 0.4(2) & 0.009344(17) & 1.6 & 0.009381(56) \\ \hline
\end{tabular}
\caption{Results of the best fits of our $\SU(3)$ numerical data according to eq.~\eqref{eq:BNG_groundstate_Nt11} up to order $1/N_t^7$ and to order $1/N_t^9$ (upper table), and also up to order $1/N_t^{11}$ (lower table). In the last case, we also report (in the rightmost column) the values of $\sigma_0 a^2$ interpolated from the data in ref.~\cite{Teper:1998te}. The range of the fits in $N_t$ is given by $\left[ N_{t,\mathrm{min}}, N_{t,\mathrm{max}} \right]$.}
\label{tab:BNG_SU3}
\end{table}

Including only the correction of order $N_t^{-7}$ in eq.~\eqref{eq:BNG_groundstate_Nt11} we are not able to fit the data points from the simulations that are closest to critical temperature, i.e., those for $N_t = 1 / (a \, T_c) + 1/2$, with the coarsest lattice spacing being the only exception. On the other hand, those points can be fitted by our model if one also includes the $N_t^{-9}$ correction, which, as can be seen from eq.~\eqref{eq:BNG_groundstate_Nt11}, does not require any additional free parameter, or including also $N_t^{-11}$ corrections (which, instead, require the further free parameter $k_5$). In all cases we truncated the series expansion of the underlying Nambu--Got{\=o} contribution (see eq.~\eqref{eq:NG_taylor}) to the term that matches the finest correction included: either $1/N_t^7$, $1/N_t^9$, or $1/N_t^{11}$. The values we obtained for the string tension at zero temperature perfectly agree with the determinations from ref.~\cite{Teper:1998te}, that we report in table~\ref{tab:BNG_SU3} for comparison. 

\begin{figure}[H]
\centering
\includegraphics[width=1\textwidth]{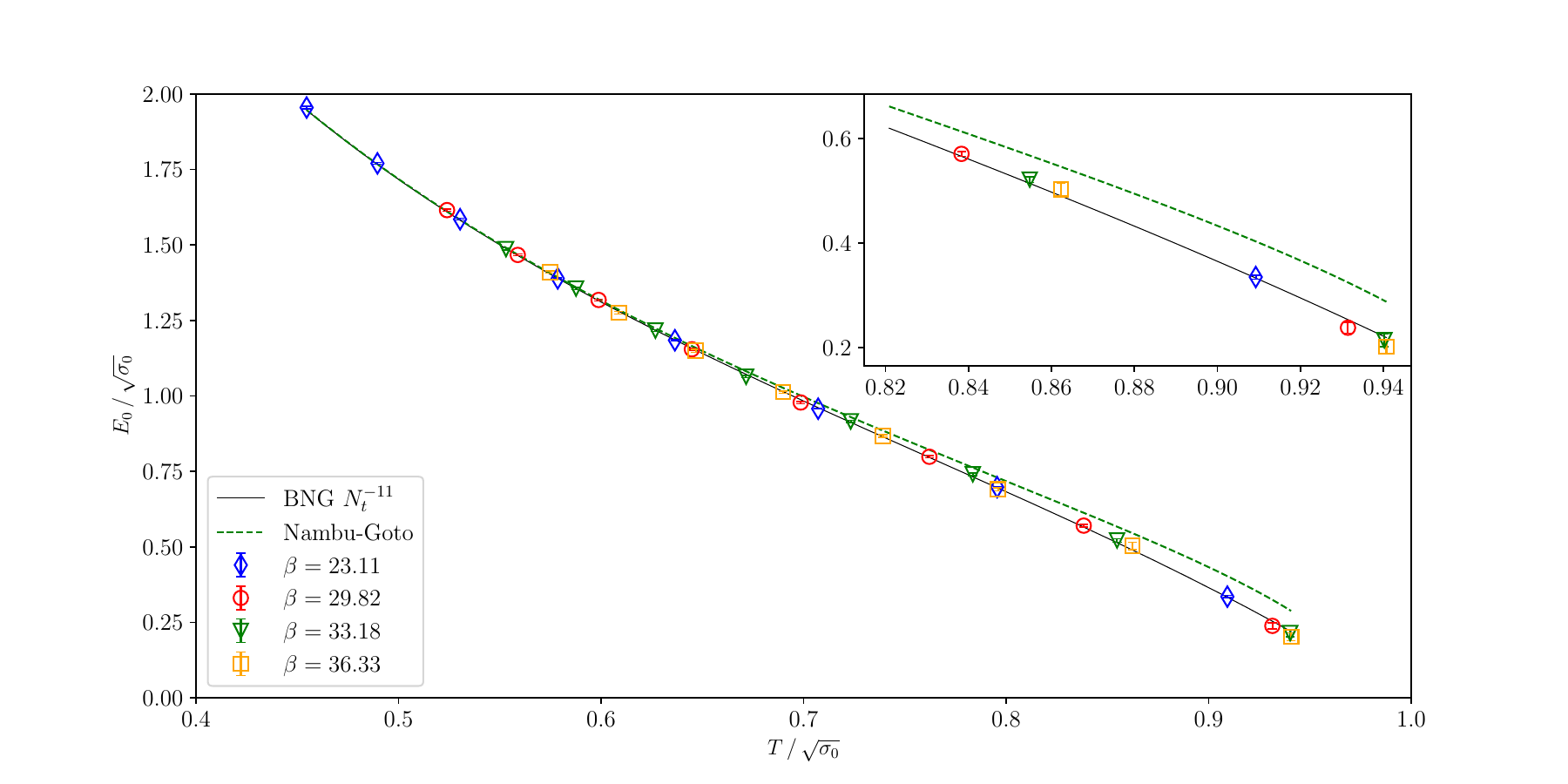}
\caption{Combined best fits of the $\SU(3)$ ground state energy $E_0$ for different values of $\beta$, according to eq.~\eqref{eq:BNG_groundstate_Nt11} including all terms up to $1/N_t^{11}$. The data are shown in units of $\sqrt{\sigma_0}$ on both axes. In the zoomed inset we show the closest points to the critical temperature where the discrepancy between our data and the NG prediction is most visible.}
\label{fig:combined_fit_SU3_Nt11}
\end{figure}

Considering all cases separately, the values of $k_4$ at different lattice spacings are compatible within their errors: this is a significant consistency check, suggesting that the scale dependence of this coefficient has already been taken into account by the $1/\sigma_0^3$ and $1/\sigma_0^4$ normalizations in the respective terms in eq.~\eqref{eq:BNG_groundstate_Nt11}. Given the absence of any clear trend when the lattice spacing is made finer, we attempt a combined fit across all values of $\beta$ fixing the same value of $k_4$ for all data sets, leaving the values of $\sigma_0 a^2$ as the remaining four independent free parameters. We perform the combined fit also for the correction up to the $N_t^{-11}$ term, which includes the parameter $k_5$. The values obtained from this procedure are presented in table~\ref{tab:BNG_SU3_combined}. Combined fits up to $N_t^{-9}$ terms and up to $N_t^{-11}$ terms are shown in fig.~\ref{fig:combined_fit_SU3_Nt9} and fig.~\ref{fig:combined_fit_SU3_Nt11}, respectively.

\begin{figure}
\centering
\includegraphics[width=0.6\textwidth]{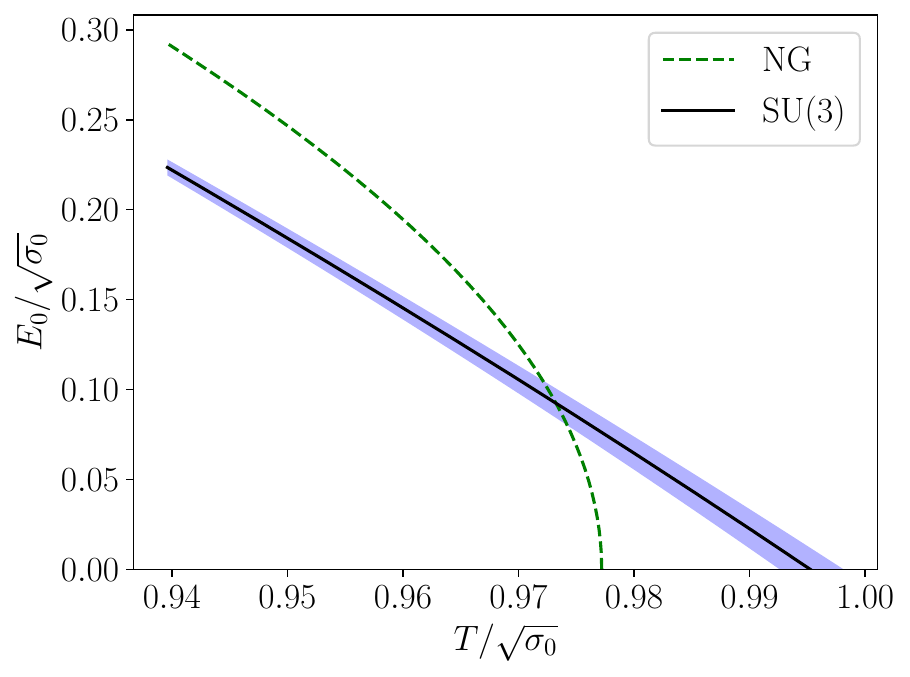}
\caption{Detail of the dependence of the ground state energy $E_0$ on the temperature at $T$ very close to $T_c$ for the $\SU(3)$ gauge theory, using our best fit parameters for the extrapolation (solid line with statistical confidence band). It is evident how the line crosses the NG one before the critical temperature and reaches the $E_0 = 0$ axis at a temperature which is compatible with the critical temperature of the theory.}
\label{fig:NG_vs_SU3}
\end{figure}

Figure~\ref{fig:combined_fit_SU3_Nt11} makes it evident that the data points systematically lie below the NG curve, hence the negative value of $k_4$ in the fit. However, it is possible to extrapolate the value of the ground state energy $E_0$ to higher temperatures, using the model truncated at order $N_t^{-11}$ and the values of $k_4$ and $k_5$ in the last row of table~\ref{tab:BNG_SU3_combined}. By this procedure, we would find, as shown in fig.~\ref{fig:NG_vs_SU3}, that at a temperature around $T = 0.973(5) \sqrt{\sigma_0} \lesssim T_{c,\mathrm{NG}}$ the model line would cross the NG one, and that the former would reach the axes $E_0$ (which we expect in correspondence of the second order phase transition) at $T_{E_0 = 0} = 0.995(5) \sqrt{\sigma_0}$, which is compatible with the critical value $T_c = 0.9890(31)\sqrt{\sigma_0}$ found in ref.~\cite{Liddle:2008kk}.

\begin{table}[H]
\centering
\begin{tabular}{|l|l|l|l|}
\hline
 & $k_4$ & $k_5$ & $\chi^2/N_{\mathrm{d.o.f.}}$ \\ \hline
up to $N_t^{-7}$ terms & -0.223(6) & -- & 1.5 \\ \hline
up to $N_t^{-9}$ terms & -0.075(2) & -- & 2 \\ \hline
up to $N_t^{-11}$ terms & -0.102(11) & 0.45(8) & 1.9 \\ \hline
\end{tabular}
\caption{Results of the best fits of our $\SU(3)$ numerical data according to eq.~\eqref{eq:BNG_groundstate_Nt11} combining all available lattice spacings.}
\label{tab:BNG_SU3_combined}
\end{table}

\begin{figure}[H]
\centering
\includegraphics[width=0.6\textwidth]{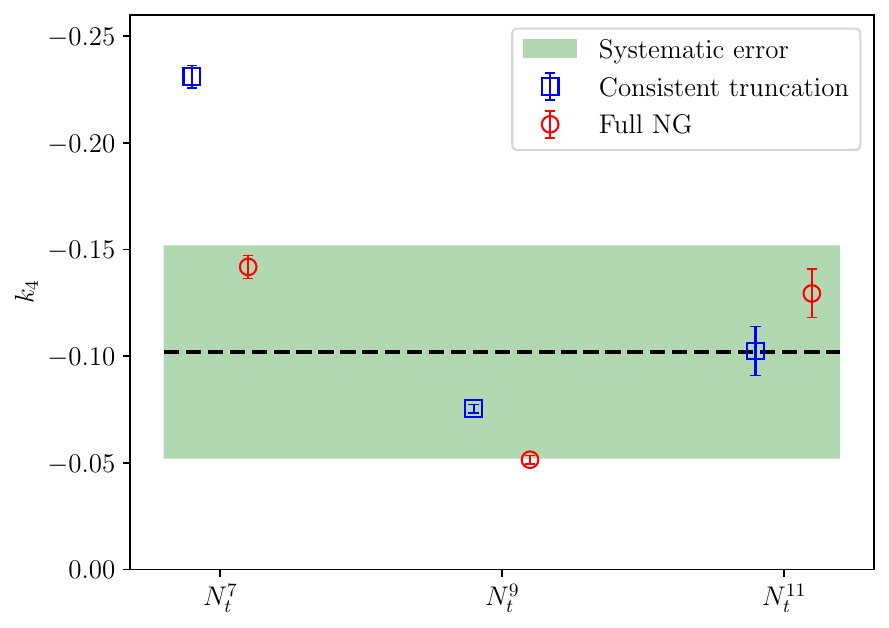}
\caption{Values of the $k_4$ for the $\SU(3)$ theory obtained through various combined fits, considering corrections beyond the Nambu--Got{\=o} string up to orders $1/N_t^7$, $1/N_t^9$ and $1/N_t^{11}$, both truncating the NG baseline to the order consistent with the finest correction (blue circles) and keeping all the NG orders (red squares). We also show our final estimate, with the systematic confidence band (green band).}
\label{fig:k4_syst_SU3}
\end{figure}

From the inset of fig.~\ref{fig:combined_fit_SU3_Nt11} the result of the combined fit is striking: with just two more parameters it improves considerably upon the NG fit at higher temperatures. However, the value of the $k_4$ coefficient from different combined fits fluctuates considerably, when different orders are taken into account in the correction, see fig.~\ref{fig:k4_syst_SU3}, indicating that the systematic error due to the truncation of the series is the most relevant source of uncertainty in our results for $k_4$. 

In order to evaluate this systematic uncertainty, for each order we repeated the fit without truncating the NG series and compared it to the results reported above, obtained by truncating the NG contribution to the ground state to the order corresponding to the finest BNG correction. From fig.~\ref{fig:k4_syst_SU3} we can observe a better agreement between these two prescriptions when higher-order BNG corrections are included. As our final central value, we choose the result obtained truncating consistently both the NG series and the corrections at the $1/N_t^{11}$ order, which we assume to be the least affected by systematic effects. This assumption seems to be supported by the relatively mild variation of $k_4$ between the $1/N_t^9$ order and the $1/N_t^{11}$ one under the consistent truncation prescription. Furthermore, we also note that the $\chi^2$ value is typically smaller for the fits performed under the consistent truncation prescription. The systematic error is chosen in order to include the $k_4$ values obtained truncating the BNG corrections at the $1/N_t^9$ and $1/N_t^{11}$ orders under both prescriptions. We assume, instead, that the value obtained truncating consistently the NG and the corrections series to the $1/N_t^7$ is affected by a very strong systematic effect due to the order of truncation.\footnote{Note, however, that our systematic error includes the points at the $1/N_t^7$ order obtained truncating the correction series, but not the NG series.}

We quote as a final result for the $\SU(3)$ theory
\begin{align}
k_4 = -0.102 (11) [50],
\end{align}
where the number in round parentheses represents the statistical error, and the one in square brackets the systematic error.

Clearly, also the value of $k_5$ is affected by a similar systematic uncertainty: we estimated it performing a fit without truncating the Nambu--Got{\=o} prediction for the ground state energy, obtaining
\begin{align}
 k_4 = -0.129(11), \; \; k_5 = 0.719(81), \; \; \chi^2 / N_{\mathrm{d.o.f.}} = 1.5.
\end{align}
As for $k_4$, the final result we report here the value obtained from the fit up to order $N_t^{-11}$, while the systematic error is calculated as the difference with the value obtained from the fit up to order $N_t^{-11}$ without the Taylor expansion of the Nambu--Got{\=o} prediction. The final result reads
\begin{align}
 k_5 = 0.45 (8) [25].
\end{align}

\subsection{Corrections beyond the Nambu--Got{\=o} string in $\SU(6)$ Yang--Mills theory}
\label{sec:su6res}

The study of the results for the $\SU(6)$ Yang--Mills theory follows closely the procedure applied in the $\SU(3)$ case, both regarding the numerical simulations and the analysis. Also in this case, we chose the values of $\beta$ for our simulations such that $1 / (a \, T_c)=6.5, 8.5, 9.5$: see table~\ref{tab:su6tableconf} for an overview of the numerical setup.

\begin{table}[H]
\centering
\begin{subtable}[b]{0.45\textwidth}
\centering
\begin{tabular}{|l|l|l|l|l|}
\hline
$\beta$ & $N_t$ & $N_s$ & $T/T_c$ & $n_{\mathrm{conf}}$ \\ \hline
\multirow{8}{*}{92} & 7 & 96 & 0.93 & $1.2\times10^5$ \\ \cline{2-5} 
 & 8 & 96 & 0.81 & $1.2\times10^5$ \\ \cline{2-5} 
 & 9 & 96 & 0.72 & $1.2\times10^5$ \\ \cline{2-5} 
 & 10 & 96 & 0.65 & $1.2\times10^5$ \\ \cline{2-5} 
 & 11 & 96 & 0.59 & $1.2\times10^5$ \\ \cline{2-5} 
 & 12 & 96 & 0.54 & $1.2\times10^5$ \\ \cline{2-5} 
 & 13 & 96 & 0.50 & $1.2\times10^5$ \\ \cline{2-5} 
 & 14 & 96 & 0.47 & $1.2\times10^5$ \\ \hline
 \end{tabular}
 \caption{$\beta=92$, $1 / (a \, T_c) = 6.5$.}
 \label{su6tableconf1}
 \end{subtable}
 \hfill
 \begin{subtable}[b]{0.45\textwidth}
 \centering
 \begin{tabular}{|l|l|l|l|l|}
 \hline
 $\beta$ & $N_t$ & $N_s$ & $T/T_c$ & $n_{\mathrm{conf}}$ \\ \hline
 \multirow{8}{*}{118} & 9 & 96 & 0.95 & $1.2\times10^5$ \\ \cline{2-5} 
 & 10 & 96 & 0.85 & $1.2\times10^5$ \\ \cline{2-5} 
 & 11 & 96 & 0.77 & $1.2\times10^5$ \\ \cline{2-5} 
 & 12 & 96 & 0.71 & $1.2\times10^5$ \\ \cline{2-5} 
 & 13 & 96 & 0.66 & $1.2\times10^5$ \\ \cline{2-5} 
 & 14 & 96 & 0.61 & $1.2\times10^5$ \\ \cline{2-5} 
 & 15 & 96 & 0.57 & $1.2\times10^5$ \\ \cline{2-5} 
 & 16 & 96 & 0.53 & $1.2\times10^5$ \\ \hline
 \end{tabular}
 \caption{$\beta=118$, $1 / (a \, T_c) = 8.5$.}
 \label{su6tableconf2}
 \end{subtable}
 
 \vspace{0.5cm}
 
 \begin{subtable}[b]{0.9\textwidth}
 \centering
 \begin{tabular}{|l|l|l|l|l|}
 \hline
 $\beta$ & $N_t$ & $N_s$ & $T/T_c$ & $n_{\mathrm{conf}}$ \\ \hline
 \multirow{8}{*}{131} & 10 & 160 & 0.95 & $1.2\times10^5$ \\ \cline{2-5} 
 & 11 & 96 & 0.87 & $1.2\times10^5$ \\ \cline{2-5} 
 & 12 & 96 & 0.79 & $1.2\times10^5$ \\ \cline{2-5} 
 & 13 & 96 & 0.73 & $1.2\times10^5$ \\ \cline{2-5} 
 & 14 & 96 & 0.68 & $1.2\times10^5$ \\ \cline{2-5} 
 & 15 & 96 & 0.63 & $1.2\times10^5$ \\ \cline{2-5} 
 & 16 & 96 & 0.60 & $1.2\times10^5$ \\ \cline{2-5} 
 & 17 & 96 & 0.56 & $1.2\times10^5$ \\ \hline
 \end{tabular}
 \caption{$\beta=131$, $1 / (a \, T_c) = 9.5$.}
 \label{su6tableconf3}
 \end{subtable}
 
 \caption{Details of our simulations of the $\SU(6)$ gauge theory at different values of the lattice spacing.}
 \label{tab:su6tableconf}
\end{table}

Also in this case, we successfully fitted our data for $R > \xi_l$ with eq.~\eqref{eq:high_dist_corr}, see tables~\ref{tab:su6tableres1}, \ref{tab:su6tableres2}, and~\ref{tab:su6tableres3}. Again, we performed the best fits of the ground state energy $E_0$ to eq.~\eqref{eq:BNG_groundstate_Nt11}, for the three different lattice spacings considered: the results are reported in table~\ref{tab:BNG_SU6}. We found values of the zero-temperature string tension consistent with those reported in the literature~\cite{Athenodorou:2016ebg}.

\begin{table}[H]
\centering
\begin{tabular}{|l|l|l|l|l|l|l|l|}
\hline
&$\beta$ & $N_{t,\mathrm{min}}$ & $N_{t,\mathrm{max}}$ & $k_4$ & $\sigma_0 a^2$ & $\chi^2/N_{\mathrm{d.o.f.}}$ \\ \hline
\multirow{ 3}{*}{up to $N_t^{-7}$} &92 & 7 & 14 & -0.247(15) & 0.02839(3) & 0.7 \\ \cline{2-7}
&118 & 9 & 16 & -0.23(2) & 0.01639(3) & 0.6 \\ \cline{2-7}
&131 & 10 & 17 & -0.244(17) & 0.013168(19) & 0.7 \\ \hline\hline

\multirow{ 3}{*}{up to $N_t^{-9}$} &92 & 7 & 14 & -0.086(6) & 0.02840(4) & 1.5 \\ \cline{2-7}
&118 & 9 & 16 & -0.080(8) & 0.01636(3) & 0.6 \\ \cline{2-7}
&131 & 10 & 17 & -0.083(6) & 0.013146(18) & 0.9 \\ \hline
\end{tabular}\\
\vspace{0.5cm}
\begin{tabular}{|l|l|l|l|l|l|l|l||l|}
\hline
&$\beta$ & $N_{t,\mathrm{min}}$ & $N_{t,\mathrm{max}}$ & $k_4$ & $k_5$ & $\sigma_0 a^2$ & $\chi^2/N_{\mathrm{d.o.f.}}$ & $\sigma_0 a^2$ from ref.~\cite{Athenodorou:2016ebg} \\ \hline
\multirow{ 3}{*}{up to $N_t^{-11}$} & 92 & 7 & 14 & -0.20(5)& 1.20(4) & 0.02839(3) & 0.5 & 0.02842(5) \\ \cline{2-9}
& 118 & 9 & 16 & -0.16(5) & 0.9(4) & 0.01640(4) & 0.7 & 0.016302(48) \\ \cline{2-9}
& 131 & 10 & 17 & -0.16(6) & 0.9(4)& 0.01317(3) & 0.8 & 0.013005(43) \\ \hline
\end{tabular}
\caption{Results of the best fits of our $\SU(6)$ numerical data according to eq.~\eqref{eq:BNG_groundstate_Nt11} up to order $1/N_t^7$ and to order $1/N_t^9$ (upper table), and also up to order $1/N_t^{11}$ (lower table). In the last case, in the rightmost column we report the values of $\sigma_0 a^2$ interpolated from the data in ref.~\cite{Athenodorou:2016ebg}.}
\label{tab:BNG_SU6}
\end{table}

\begin{figure}[H]
 \centering 
 \includegraphics[width=1\textwidth]{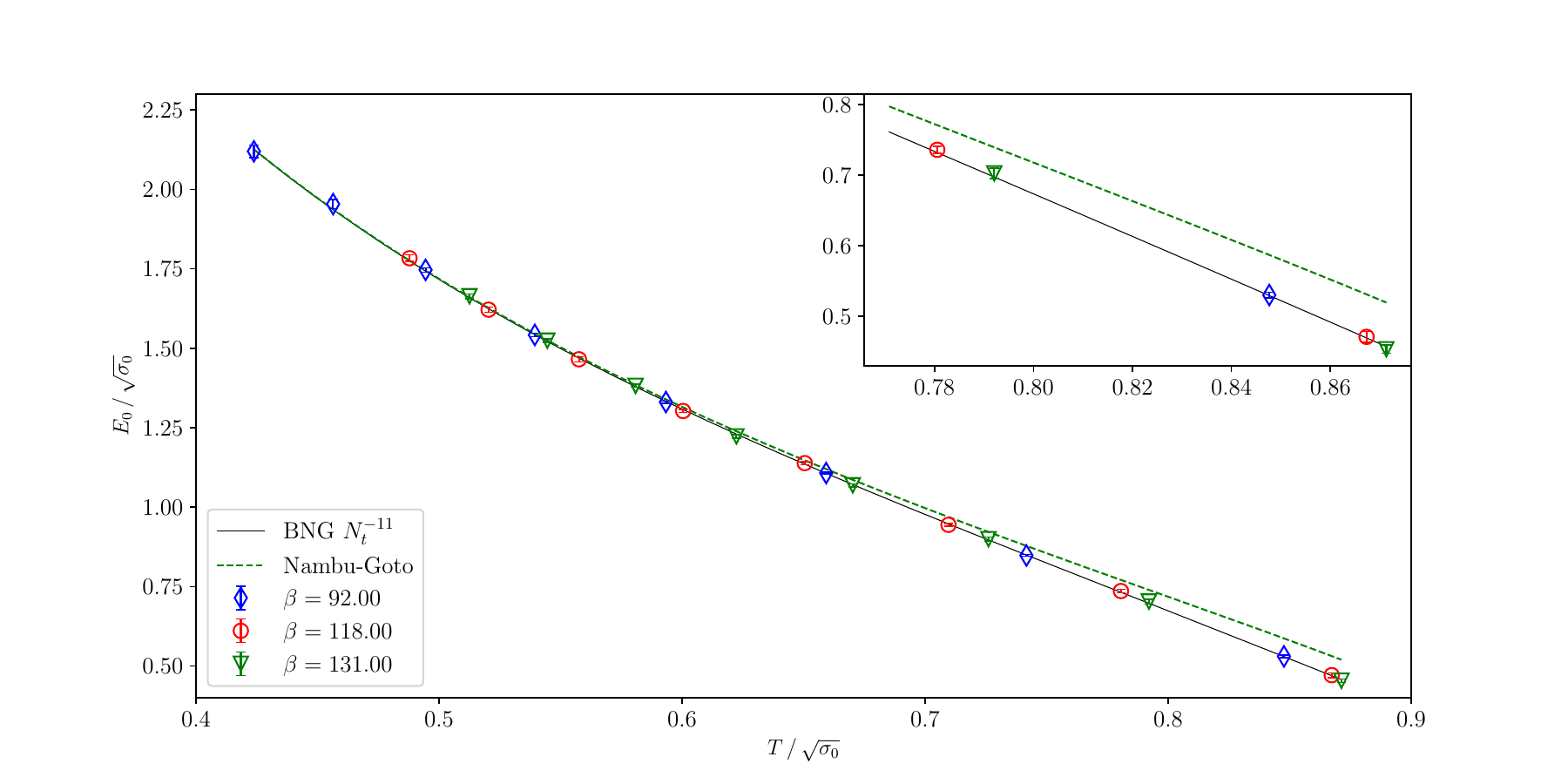}
 \caption{Combined best fits of the ground-state energy $E_0$ in the $\SU(6)$ theory, for different values of $\beta$, according to eq.~\eqref{eq:BNG_groundstate_Nt11} including all terms up to $1/N_t^{11}$. The data points are shown in units of $\sqrt{\sigma_0}$ on both axes. The zoomed inset shows the closest points to the critical temperature, where the discrepancy between our data and the NG prediction is largest.}
 \label{fig:combined_fit_SU6_Nt11}
\end{figure}

Analogously to our results for the $\SU(3)$ theory, the values of $k_4$ are perfectly compatible with each other for different values of the lattice spacing, allowing a combined fit as the one performed in subsection~\ref{sec:su3res}. The values of $k_4$ obtained from the combined fits are reported in table~\ref{tab:BNG_SU6_combined} and displayed in fig.~\ref{fig:combined_fit_SU6_Nt9} and in fig.~\ref{fig:combined_fit_SU6_Nt11}.

\begin{table}[H]
\centering
\begin{tabular}{|l|l|l|l|}
\hline
 & $k_4$ & $k_5$ & $\chi^2/N_{\mathrm{d.o.f.}}$ \\ \hline
Up $N_t^{-7}$ terms & -0.242(10) & -- & 0.6\\ \hline
Up $N_t^{-9}$ terms & -0.084(4) & -- & 1.0\\ \hline
Up $N_t^{-11}$ terms & -0.173(30) & 0.98(23) & 1.0\\ \hline
\end{tabular}
\caption{Results of the best fits of our $\SU(6)$ numerical data according to eq.~\eqref{eq:BNG_groundstate_Nt11}, combining all available lattice spacings.}
\label{tab:BNG_SU6_combined}
\end{table}

\begin{figure}[H]
 \centering
 \includegraphics[width=0.6\textwidth]{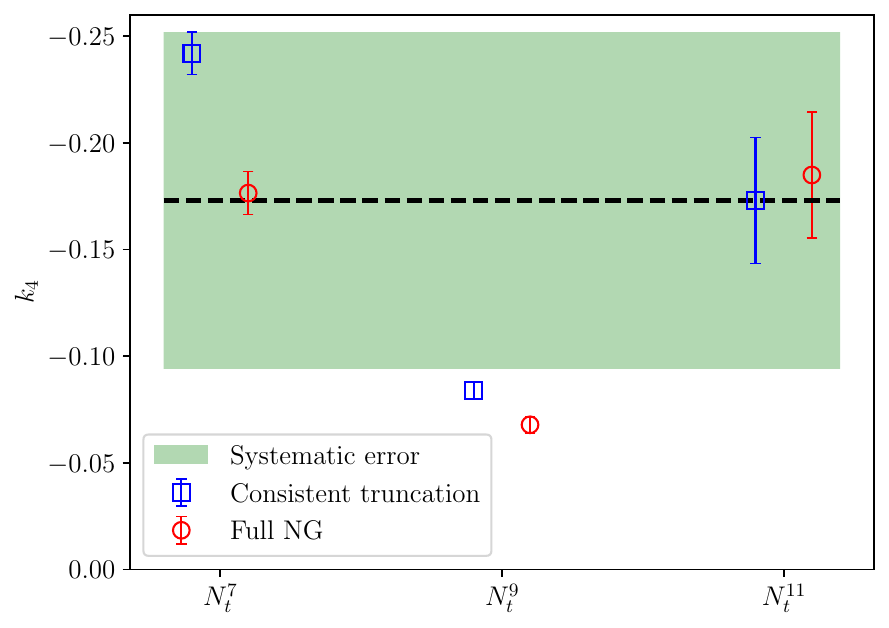}
 \caption{$k_4$ values for the $\SU(6)$ theory, as obtained from various fits, considering corrections of order $1/N_t^7$, $1/N_t^9$ and $1/N_t^{11}$, both truncating the NG baseline to the order consistent with the finest correction (blue circles) and keeping the complete Nambu--Got{\=o} prediction (red squares). Our final estimate, with the systematic confidence band, is shown in green.}
 \label{fig:k4_syst_SU6}
\end{figure}

Like for the $\SU(3)$ Yang--Mills theory, a reliable determination of the systematic error due to the order of the corrections is crucial also for the results of the $\SU(6)$ theory. To this purpose, we studied the values of $k_4$ obtained with a variety of different truncation prescriptions: see fig.~\ref{fig:k4_syst_SU6} for an overview. The central value of our final result is, again, the result of the fit where we truncated both the NG baseline and the correction series to the $1/N_t^{11}$ order. This time, for the systematic error, we consider the semi-dispersion between the values obtained with the same prescription, truncating the correction to the $1/N_t^9$ and $1/N_t^7$ order. We quote as final results
\begin{align}
k_4 = -0.173(30)[79].
\end{align}
where the first uncertainty is the statistical error, and the second is the systematic one.

Concerning $k_5$, we repeated the fits including all the orders of the underlying NG contribution, obtaining the following result:
\begin{align}
 k_4 = -0.185(30), \; \; k_5 = 1.11(23), \; \; \chi^2/N_{\mathrm{d.o.f.}} = 0.54.
\end{align}
Since the discrepancy in the values of $k_5$ is smaller than a standard deviation, we quote the following final value:
\begin{align}
k_5 = 0.98(23)[15].
\end{align}

\subsection{Comparing with results from the $\SU(2)$ theory}

\begin{figure}[H]
 \centering 
 \includegraphics[width=1\textwidth]{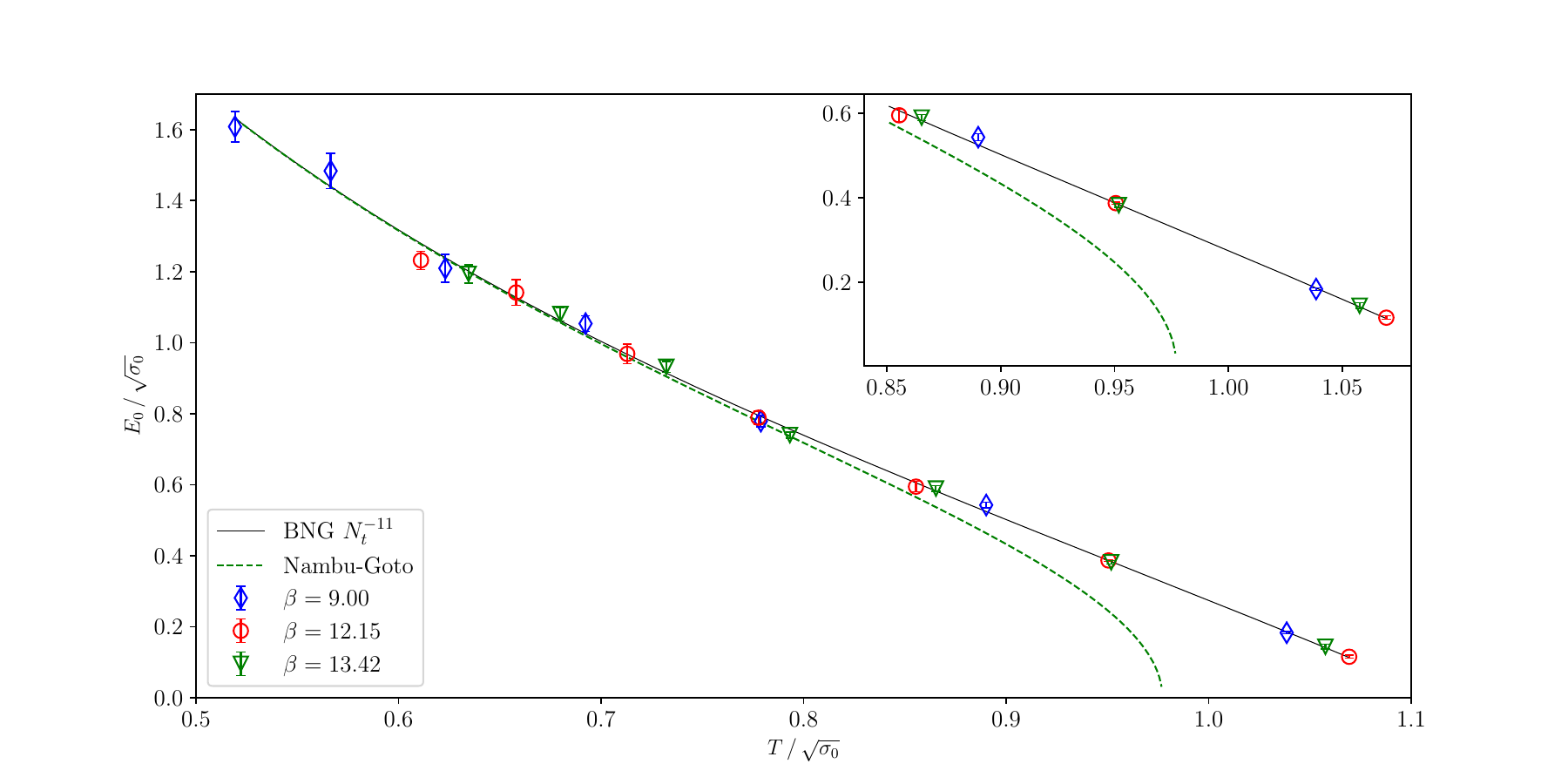}
 \caption{Combined best fits of the ground state energy $E_0$ up to $1/N_t^1$ terms at different $\beta$ according to \eqref{eq:BNG_groundstate_Nt11}, for the $\SU(2)$ theory, with data from ref.~\cite{Caristo:2021tbk}. The quantities on both axes are in terms of the square root of the zero-temperature string tension $\sigma_0$. The inset show the region where the deviation of the lattice data with respect to the NG prediction becomes largest.}
 \label{fig:combined_fit_SU2_Nt11}
\end{figure}

The next step in our analysis consists in comparing our results with the values previously obtained for $\SU(2)$ Yang--Mills theory in ref.~\cite{Caristo:2021tbk}. In order to follow a similar procedure to the $\SU(3)$ and $\SU(6)$ cases, rather than using the value of $k_4$ quoted in ref.~\cite{Caristo:2021tbk} (which was obtained with a weighted average), we start from the values of the ground state $E_0$ taken from ref.~\cite[tables 7, 8, and 9]{Caristo:2021tbk}) and repeat the combined fit for all lattice spacings that we used in the previous sections. The results of this analysis are reported in table~\ref{tab:BNG_SU2_combined} and displayed in fig.~\ref{fig:combined_fit_SU2_Nt9} and in fig.~\ref{fig:combined_fit_SU2_Nt11}.

\begin{table}[H]
\centering
\begin{tabular}{|l|l|l|l|}
\hline
 & $k_4$ & $k_5$ & $\chi^2/N_{\mathrm{d.o.f.}}$ \\ \hline
Up $N_t^{-7}$ terms & 0.050(3) & -- & 1.3 \\ \hline
Up $N_t^{-9}$ terms & 0.0258(10) & -- & 1.3 \\ \hline
Up $N_t^{-11}$ terms & 0.0386(95) & -0.123(52) & 1.3 \\ \hline
\end{tabular}
\caption{Results of the best fits of $\SU(2)$ numerical data from ref.~\cite{Caristo:2021tbk} according to eq.~\eqref{eq:BNG_groundstate_Nt11}, combining the data at all available lattice spacings.}
\label{tab:BNG_SU2_combined}
\end{table}

Note that in this case, given the positive value of $k_4$, it is actually impossible to perform the fit without truncating the NG series. Keeping the square root, indeed, the data points that are closest to the critical temperature would have an imaginary contribution from the NG term, due to the fact that in $2+1$ dimensions the critical temperature of the $\SU(2)$ Yang--Mills theory is larger than $T_{c,\mathrm{NG}}$. This is clearly visible in the high-temperature region in fig.~\ref{fig:combined_fit_SU2_Nt11}, where the NG prediction cannot be extended beyond $T_{c,\mathrm{NG}} / \sqrt{\sigma_0}= \sqrt{3 / \pi} = 0.977\dots$, see eq.~\eqref{eq:TCNG}. For this reason, we always truncate consistently the NG series and the corrections to the same order and estimate the systematic error as half of the difference between the values obtained from truncating at the $N_t^7$ and the $N_t^9$ order. In this case, our final result is:
\begin{align}
 k_4 = 0.0386(95) [121].
\end{align}
For the same reason discussed above, in this case we do not assign a systematic uncertainty to the $k_5$ value in table~\ref{tab:BNG_SU2_combined}. Note, however, that the relative statistical uncertainty on this result is relatively large, and probably dominates over the systematic one in the total error budget. 

\begin{figure}[h]
 \centering
 \includegraphics[width=0.6\textwidth]{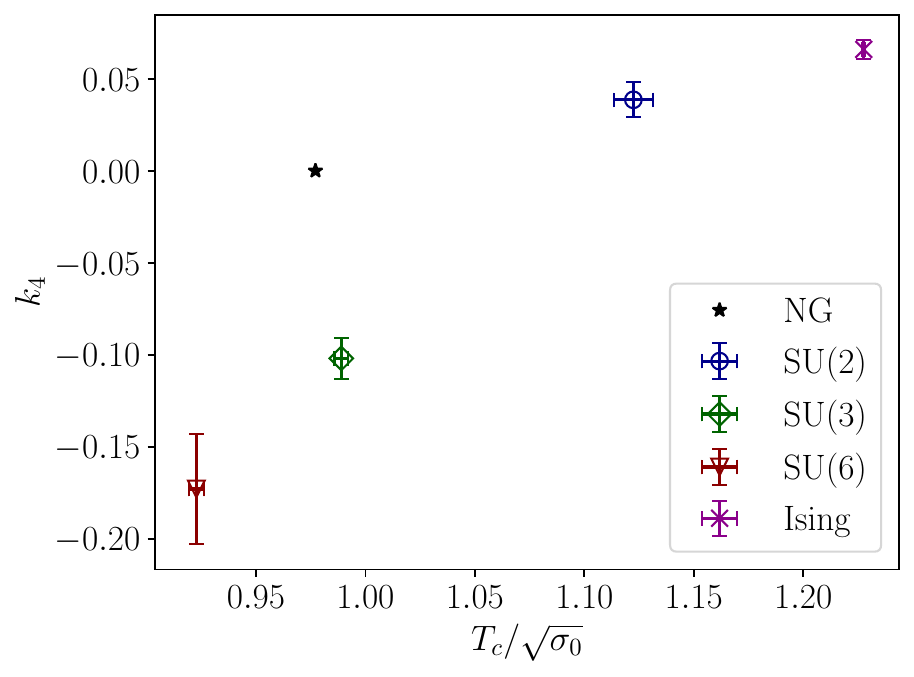}
 \caption{$k_4$ values for different confining gauge theories in three spacetime dimensions, plotted as a function of the critical temperature in units of the square root of the zero-temperature string tension $\sigma_0$, for the $\SU(N)$ gauge theories considered in this work and for the $\Z_2$ gauge theory that was studied in ref.~\cite{Baffigo:2023rin}.}
 \label{fig:k4_vs_Tc}
\end{figure}

Finally, having now determined the values of the $k_4$ coefficient for $\SU(N)$ Yang--Mills theories in three dimensions for $N=2$, $3$, and $6$ color charges, it is interesting to compare them together, and with the one obtained for the $\Z_2$ gauge theory in ref.~\cite{Baffigo:2023rin}: we plot these results as a function of the critical temperature in units of the square root of the string tension in fig.~\ref{fig:k4_vs_Tc}. First of all, we note that the $k_4$ coefficient appears to be decreasing with the ``size'' of the gauge group:\footnote{The \emph{dimension} of $\SU(N)$ Lie groups is $N^2-1$; obviously, this notion cannot be directly compared with the \emph{order} of finite groups, such as $\Z_N$, which is $N$, even though both concepts are related to the number of microscopic internal degrees of freedom of the corresponding gauge theory.} in particular, $k_4$ is positive for the $\Z_2$ and $\SU(2)$ gauge theories, while it is negative for the $\SU(3)$ and $\SU(6)$. Even though the numerical values are close to each other, the statistical and systematic uncertainties that we estimated seem to suggest that for the $\SU(6)$ theory $k_4$ is larger in magnitude (hence more negative).

Moreover, the results in fig.~\ref{fig:k4_vs_Tc} are also consistent with the idea that the critical temperature of each of these theories correlates with the $k_4$ correction to Nambu--Got{\=o} in the effective string action. However, this relation does not seem to be trivial: in the $\SU(3)$ case, for example, the relative difference between $T_{c,\SU(3)}$ and $T_{c,\mathrm{NG}}$ is less than $2 \%$, and thus one may have na{\"{\i}}vely expected a much smaller absolute value for $k_4$, which does not seem to be the case.

\subsection{Comparison with bootstrap constraints}

\begin{figure}[h]
\centering
\includegraphics[width=0.6\textwidth]{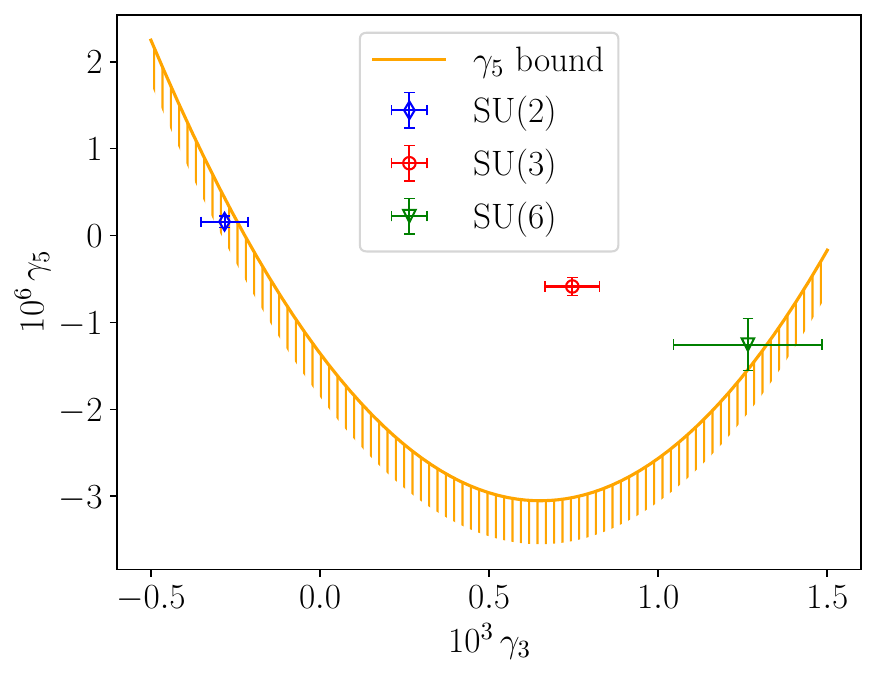}
\caption{Values of $\gamma_3$ and $\gamma_5$ for the $\SU(N)$ theories studied in this work. The error bars represent only the statistical uncertainty associated to the numerical value. The solid line denotes the lower bound on $\gamma_5$ from the bootstrap analysis (the hatched side is the forbidden region), while the bound on $\gamma_3$ ($\gamma_3 > -0.0013\dots)$ lies outside of the region shown in the plot, on its left.}
\label{fig:gamma_bound}
\end{figure}

As the last step in the analysis of the corrections of the Nambu--Got{\=o} contribution to $E_0$, we compare our results for the $k_4$ and $k_5$ parameters with the bounds found from the S-matrix bootstrap analysis performed in ref.~\cite{EliasMiro:2019kyf}. Using eq.~\eqref{eq:ground_state_smatrix}, $k_{4}$ can be expressed in terms of $\gamma_{3}$, while $k_{5}$ can be related to $\gamma_{3,5}$:
\begin{align}
\gamma_3 = - \frac{225}{32 \pi^6} k_4, \;\;\;\;\; \gamma_5 = - \frac{3969}{32768 \pi^{10}} k_5.
\end{align}
For convenience, we summarize these values in table~\ref{tab:BNG_SU236_gamma} and plot them in fig.~\ref{fig:gamma_bound}.

\begin{table}[H]
\centering
\begin{tabular}{|c|c|c|}
\hline
 & $\gamma_3 \times 10^3$ & $\gamma_5 \times 10^6$ \\
\hline
$\SU(2)$ & $-0.282 (70) [89] $ & $ 0.159 (66) $ \\
\hline
$\SU(3)$ & $ 0.746 (80) [365]$ & $-0.58 (11) [32]$ \\
\hline
$\SU(6)$ & $ 1.26 (22) [58] $ & $-1.25 (30) [19]$ \\
\hline
\end{tabular}
\caption{Results of the coefficients $\gamma_3$ and $\gamma_5$ from eq.~\eqref{eq:ground_state_smatrix} for various gauge theories.}
\label{tab:BNG_SU236_gamma}
\end{table}

Firstly, all values of $\gamma_3$ are well inside the bound $\gamma_3 > - \frac{1}{768} \simeq -0.0013$. Furthermore, the bound on $\gamma_5$, which is denoted by the solid yellow line (with the hatched region being the excluded one), depends on the one on $\gamma_3$ according to eq.~\eqref{eq:bound_bootstrap}. The values $\SU(3)$ and $\SU(6)$ from our analysis are in the allowed region; the result for $\SU(2)$, in contrast, lies outside of it, but not significantly, and considering the combination of statistical and systematic uncertainties it may be compatible with the bound as well. 

The values we found for the $\SU(3)$ and $\SU(6)$ theories are slightly larger than what was predicted in refs.~\cite{Guerrieri:2024ckc,Dubovsky:2014fma}. In particular our $\SU(6)$ value is on the edge of the most constraining bound in their \emph{branon matryoshka}. To compute that bound, indeed, the inequality $\gamma_3 < 10^{-3}$ was used, which is satisfied by our result within $1.2$ statistical errors. 

\begin{figure}
    \centering
    \includegraphics[width=0.6\textwidth]{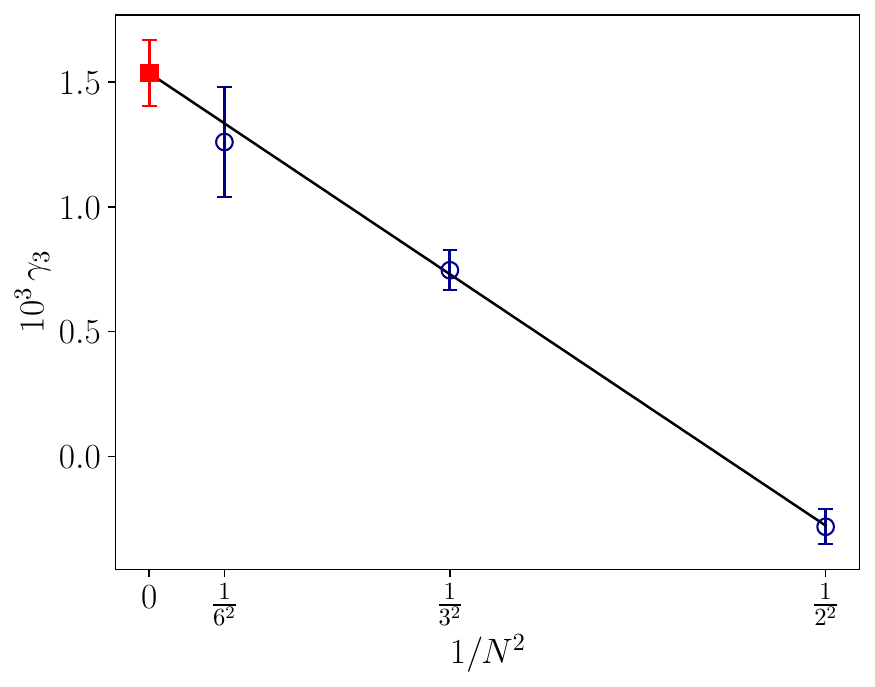}
    \caption{Large $N$ extrapolation for $\gamma_3$ using our final values obtained for $N = 2, 3$ and $6$. the solid line is the best fit assuming corrections of the form expressed in \ref{gamma3_vs_N}. Filled square represent the extrapolated value at $1 / N^2 = 0$.}
    \label{fig:largeN_gamma3}
\end{figure}

The $\gamma_3$ values that we found for the $\SU(3)$ and $\SU(6)$ theories are quite similar to each other, suggesting a weak dependence on $N$ for $N \ge 3$; this would be in line with what is generally observed for other observables in $\SU(N)$ Yang--Mills theories, both in $2+1$ and in $3+1$ dimensions~\cite{Lucini:2012gg, Panero:2012qx}
It would be interesting to estimate the value of $\gamma_3$ in the large-$N$ limit. Assuming a dependence of $\gamma_3$ on the number of color charges of the form
\begin{align}
\label{gamma3_vs_N}
\gamma_3^{(N = \infty)} + \frac{c}{N^2},
\end{align}
we performed the best fit (shown in fig.~\ref{fig:largeN_gamma3}) of our numerical data including the results from the $\SU(3)$, the $\SU(6)$, and also the $\SU(2)$ theory. The result of the fit is $\gamma_3^{(N = \infty)}  =1.54(13)\times 10^{-3}$, which is within one standard deviation from our result for the $\SU(6)$ Yang--Mills theory.

\section{Conclusions}
\label{sec:conclusions}

In this work we carried out a systematic study of the effective string corrections beyond the Nambu--Got{\=o} action, for $\SU(N)$ Yang--Mills theories in $2+1$ spacetime dimensions. One of the main goals of our analysis -- which, at least to some extent, can be visualized through fig.~\ref{fig:k4_vs_Tc} -- consisted in investigating the fine details characterizing the confining dynamics in theories based on different gauge groups. As a matter of fact, while it is known that the Nambu--Got{\=o} bosonic string is an excellent, and universal, description of the long-distance properties of confining flux tubes, this universality of the confining string can (and must) be violated by strongly suppressed, high-order terms in an expansion in inverse powers of the string tension.

Our analysis was based on a new set of high-precision results of Monte Carlo simulations of the two-point Polyakov loop correlation function, determined non-perturbatively in the lattice regularization. Following the approach used in ref.~\cite{Caristo:2021tbk} for the $\SU(2)$ gauge theory, we investigated the cases of the theories with $N=3$ and $6$ color charges in the proximity of their deconfinement phase transition at finite temperature. We extracted the ground-state energy of the effective string and we estimated the deviation from the Nambu--Got{\=o} approximation, which can be parametrized in terms of the $k_4$ and $k_5$ coefficients appearing in the expansion of the correlator around the long-string limit.

The final results quoted in sections~\ref{sec:su3res} and~\ref{sec:su6res} are reported with their statistical and systematic uncertainties; the latter, in particular, arises from both the truncation of the order of the correction and the truncation of the order of the Nambu--Got{\=o} prediction, which, as we discussed above, requires a careful treatment. For this reason we provided a conservative estimate of the systematic error, which is generally larger than the statistical one.

In addition, we also reported an improved estimate of $k_4$ and a novel determination of the $k_5$ coefficient for the $\SU(2)$ Yang--Mills theory, using the data reported in ref.~\cite{Caristo:2021tbk}, with an improved estimation of systematic effects.

Our estimates for the $k_4$ and $k_5$ coefficients can be directly translated in terms of the $\gamma_3$ and $\gamma_5$ coefficients, and are found to be in agreement with the bounds obtained from the bootstrap analysis.

In parallel, we also performed a new, high-precision test of the Svetitsky--Yaffe conjecture~\cite{Svetitsky:1982gs}, comparing the analytical solution of the short-distance spin-spin correlator for the three-state Potts model in two dimensions and our data for the $\SU(3)$ gauge theory in $2+1$ dimensions at finite temperature. The functional form predicted from the Potts model by means of conformal perturbation theory was successfully fitted to the results for the Polyakov loop correlator of the gauge theory at short distances, and the fit yields a correlation length in remarkable agreement with the one obtained from the long-range fits motivated by the EST predictions. 

The fact that two completely independent effective descriptions, which are expected to be valid in different regimes of the theory, provide consistent results in a finite range is a remarkable confirmation of the predictive power of the Svetitsky--Yaffe conjecture: the details of the short-range interactions between the effective degrees of freedom of the theory in the vicinity of a continuous deconfinement transition are completely captured by a strikingly simple spin model. As we already remarked in the introduction of the present work, this could even have phenomenological applications, in particular in the context of the QCD phase diagram at finite temperature and quark chemical potential, \emph{if} it features a critical end-point. In particular, the pattern of symmetries relevant in that case would lead to the expectation of a low-energy description in terms of the three-dimensional Ising model~\cite{Halasz:1998qr, Berges:1998rc}, and, as was already pointed out in ref.~\cite{Caselle:2019tiv}, combining the predictions from the Svetitsky--Yaffe mapping with recent, very accurate results for the Ising model from the bootstrap approach~\cite{El-Showk:2012cjh, El-Showk:2014dwa, Gliozzi:2014jsa} and with conformal perturbation theory~\cite{Zamolodchikov:1987ti, Guida:1995kc, Gaberdiel:2008fn, Caselle:2016mww, Amoretti:2017aze} one could get extremely valuable theoretical insight into a region of the QCD phase diagram which (despite interesting recent insights~\cite{Fujimoto:2023unl, Chiba:2023ftg, Moore:2023glb, Navarrete:2024zgz, Abbott:2024vhj, Kojo:2024sca}) remains inaccessible to lattice simulations~\cite{Philipsen:2005mj, deForcrand:2009zkb, Aarts:2015tyj, Gattringer:2016kco}. In this respect, a proper identification of the mapping between the QCD parameters and the thermal and magnetic deformations of the critical Ising model \cite{Caselle:2020tjz
} could then lead to particularly important phenomenological predictions for heavy-ion collisions, including for the hydrodynamic evolution of QCD matter near the critical endpoint and for experimentally accessible final hadronic yields~\cite{An:2021wof, Parotto:2018pwx, Nonaka:2004pg, Kampfer:2005nt}.

\subsection*{Acknowledgements}

We thank A.~Bulgarelli and E.~Cellini for helpful discussions. This work of has been partially supported by the Italian PRIN ``Progetti di Ricerca di Rilevante Interesse Nazionale -- Bando 2022'', prot. 2022TJFCYB, and by the ``Simons Collaboration on Confinement and QCD Strings'' funded by the Simons Foundation. The simulations were run on CINECA computers. We acknowledge support from the SFT Scientific Initiative of the Italian Nuclear Physics Institute (INFN).


\begin{appendices}
\section{Fits of numerical results} \label{sec:appendix}

\begin{table}[H]
\centering
\begin{tabular}{|l|l|l|l|l|l|}
\hline
$N_t$ & $R_{\mathrm{min}}/a$ & $R_{\mathrm{max}}/a$ & $k_l$ & $\xi_l/a$ & $\chi^2/N_{\mathrm{d.o.f.}}$ \\ \hline

7 & 19 & 79 & 0.01830(16) & 19.00(21) & 0.1 \\ \hline
8 & 11 & 79 & 0.01957(7) & 9.12(3) & 0.2 \\ \hline
9 & 8 & 47 & 0.01925(4) & 6.646(11) & 0.3 \\ \hline
10 & 6 & 47 & 0.01858(3) & 5.374(6) & 0.8 \\ \hline
11 & 6 & 47 & 0.01753(4) & 4.581(7) & 0.9 \\ \hline
12 & 6 & 47 & 0.01640(5) & 4.015(7) & 1.0 \\ \hline
13 & 6 & 47 & 0.01524(8) & 3.595(7) & 1.1 \\ \hline
14 & 6 & 47 & 0.01401(8) & 3.255(8) & 1.3 \\ \hline
\end{tabular}
 \caption{Results of the fits for different values of $N_t$ at $\beta=23.11$. }
\label{tab:su3tableres1}
\end{table}

\begin{table}[H]
\centering
\begin{tabular}{|l|l|l|l|l|l|}
\hline
$N_t$ & $R_{\mathrm{min}}/a$ & $R_{\mathrm{max}}/a$ & $k_l$ & $\xi_l/a$ & $\chi^2/N_{\mathrm{d.o.f.}}$ \\ \hline
9 & 35 & 79 & 0.0144(4) & 35.7(1.1) & 0.03 \\ \hline
10 & 15 & 79 & 0.01693(12) & 14.70(10) & 0.07 \\ \hline
11 & 11 & 47 & 0.01695(7) & 10.50(5) & 0.2 \\ \hline
12 & 9 & 47 & 0.01623(5) & 8.57(3) & 0.4 \\ \hline
13 & 8 & 47 & 0.01564(4) & 7.261(17) & 0.4 \\ \hline
14 & 7 & 47 & 0.01493(4) & 6.361(14) & 1.0 \\ \hline
15 & 7 & 47 & 0.01403(4) & 5.714(9) & 1.2 \\ \hline
16 & 7 & 47 & 0.01316(5) & 5.189(13) & 0.9 \\ \hline
\end{tabular}
 \caption{Results of the fits for different values of $N_t$ at $\beta=29.82$. }
\label{tab:su3tableres2}
\end{table}

\begin{table}[H]
\centering
\begin{tabular}{|l|l|l|l|l|l|}
\hline
$N_t$ & $R_{\mathrm{min}}/a$ & $R_{\mathrm{max}}/a$ & $k_l$ & $\xi_l/a$ & $\chi^2/N_{\mathrm{d.o.f.}}$ \\ \hline
10 & 50 & 79 & 0.01317(10) & 44(3) & 0.01 \\ \hline
11 & 19 & 79 & 0.01582(17) & 18.04(22) & 0.05 \\ \hline
12 & 13 & 47 & 0.01605(9) & 12.70(8) & 0.05 \\ \hline
13 & 11 & 47 & 0.01558(7) & 10.24(5) & 0.13 \\ \hline
14 & 10 & 47 & 0.01481(6) & 8.84(3) & 0.3 \\ \hline
15 & 9 & 47 & 0.01420(5) & 7.73(2) & 0.3 \\ \hline
16 & 8 & 47 & 0.01349(5) & 6.935(19) & 0.6 \\ \hline
17 & 8 & 47 & 0.01266(5) & 6.319(16) & 1.2 \\ \hline
\end{tabular}
 \caption{Results of the fits for different values of $N_t$ at $\beta=33.18$. }
\label{tab:su3tableres3}
\end{table}

\begin{table}[H]
\centering
\begin{tabular}{|l|l|l|l|l|l|}
\hline
$N_t$ & $R_{\mathrm{min}}/a$ & $R_{\mathrm{max}}/a$ & $k_l$ & $\xi_l/a$ & $\chi^2/N_{\mathrm{d.o.f.}}$ \\ \hline
11 & 53 & 119 & 0.0129(9) & 51(3) & 0.03 \\ \hline
12 & 21 & 79 & 0.0153(2) & 20.6(3) & 0.11 \\ \hline
13 & 16 & 47 & 0.01510(15) & 15.00(12) & 0.4 \\ \hline
14 & 13 & 47 & 0.01493(10) & 11.94(6) & 0.2 \\ \hline
15 & 11 & 47 & 0.01437(7) & 10.22(4) & 0.4 \\ \hline
16 & 10 & 47 & 0.01371(6) & 9.00(4) & 0.5 \\ \hline
17 & 9 & 47 & 0.01300(6) & 8.11(3) & 0.4 \\ \hline
18 & 8 & 47 & 0.01247(5) & 7.336(19) & 1.1 \\ \hline
\end{tabular}
 \caption{Results of the fits for different values of $N_t$ at $\beta=36.33$. }
\label{tab:su3tableres4}
\end{table}

\begin{table}[H]
\centering
\begin{tabular}{|l|l|l|l|l|l|}
\hline
$N_t$ & $R_{\mathrm{min}}/a$ & $R_{\mathrm{max}}/a$ & $k_l$ & $\xi_l/a$ & $\chi^2/N_{\mathrm{d.o.f.}}$ \\ \hline
7 & 12 & 47 & 0.00500(5) & 11.19(9) & 0.2 \\ \hline
8 & 8 & 47 & 0.00487(3) & 7.00(3) & 0.4 \\ \hline
9 & 6 & 47 & 0.004683(17) & 5.359(13) & 0.8 \\ \hline
10 & 6 & 47 & 0.00439(2) & 4.459(12) & 0.9 \\ \hline
11 & 6 & 47 & 0.004008(3) & 3.848(17) & 1.0 \\ \hline
12 & 6 & 47 & 0.00376(3) & 3.397(13) & 1.1 \\ \hline
13 & 6 & 47 & 0.00350(7) & 3.04(2) & 1.2 \\ \hline
14 & 6 & 47 & 0.00304(9) & 2.780(3) & 1.1 \\ \hline
\end{tabular}
 \caption{Results of the fits for different values of $N_t$ at $\beta=92$. }
\label{tab:su6tableres1}
\end{table}

\begin{table}[H]
\centering
\begin{tabular}{|l|l|l|l|l|l|}
\hline
$N_t$ & $R_{\mathrm{min}}/a$ & $R_{\mathrm{max}}/a$ & $k_l$ & $\xi_l/a$ & $\chi^2/N_{\mathrm{d.o.f.}}$ \\ \hline
9 & 18 & 47 & 0.004364(10) & 16.6(3) & 0.1 \\ \hline
10 & 11 & 47 & 0.00423(4) & 10.61(7) & 0.4 \\ \hline
11 & 9 & 47 & 0.00404(3) & 8.27(4) & 0.5 \\ \hline
12 & 8 & 47 & 0.00389(2) & 6.86(3) & 0.5 \\ \hline
13 & 8 & 47 & 0.00361(3) & 5.99(24) & 0.7 \\ \hline
14 & 9 & 47 & 0.00336(4) & 5.33(3) & 0.9 \\ \hline
15 & 8 & 47 & 0.00312(3) & 4.81(2) & 0.9 \\ \hline
16 & 8 & 47 & 0.00292(4) & 4.38(3) & 0.9 \\ \hline
\end{tabular}
 \caption{Results of the fits for different values of $N_t$ at $\beta=118$. }
\label{tab:su6tableres2}
\end{table}

\begin{table}[H]
\centering
\begin{tabular}{|l|l|l|l|l|l|}
\hline
$N_t$ & $R_{\mathrm{min}}/a$ & $R_{\mathrm{max}}/a$ & $k_l$ & $\xi_l/a$ & $\chi^2/N_{\mathrm{d.o.f.}}$ \\ \hline
10 & 20 & 47 & 0.00415(7) & 19.2(3) & 0.09 \\ \hline
11 & 14 & 47 & 0.00405(6) & 12.40(13) & 0.1 \\ \hline
12 & 11 & 47 & 0.00391(4) & 9.69(6) & 0.4 \\ \hline
13 & 9 & 47 & 0.00371(3) & 8.15(4) & 0.4 \\ \hline
14 & 8 & 47 & 0.00347(2) & 7.13(3) & 0.3 \\ \hline
15 & 7 & 47 & 0.003312(17) & 6.31(2) & 0.8 \\ \hline
16 & 7 & 47 & 0.003090(18) & 5.72(2) & 1.0 \\ \hline
17 & 8 & 47 & 0.00286(3) & 5.24(2) & 0.8 \\ \hline
\end{tabular}
 \caption{Results of the fits for different values of $N_t$ at $\beta=131$. }
\label{tab:su6tableres3}
\end{table}

\begin{figure}[H]
 \centering 
 \includegraphics[width=0.70\textwidth]{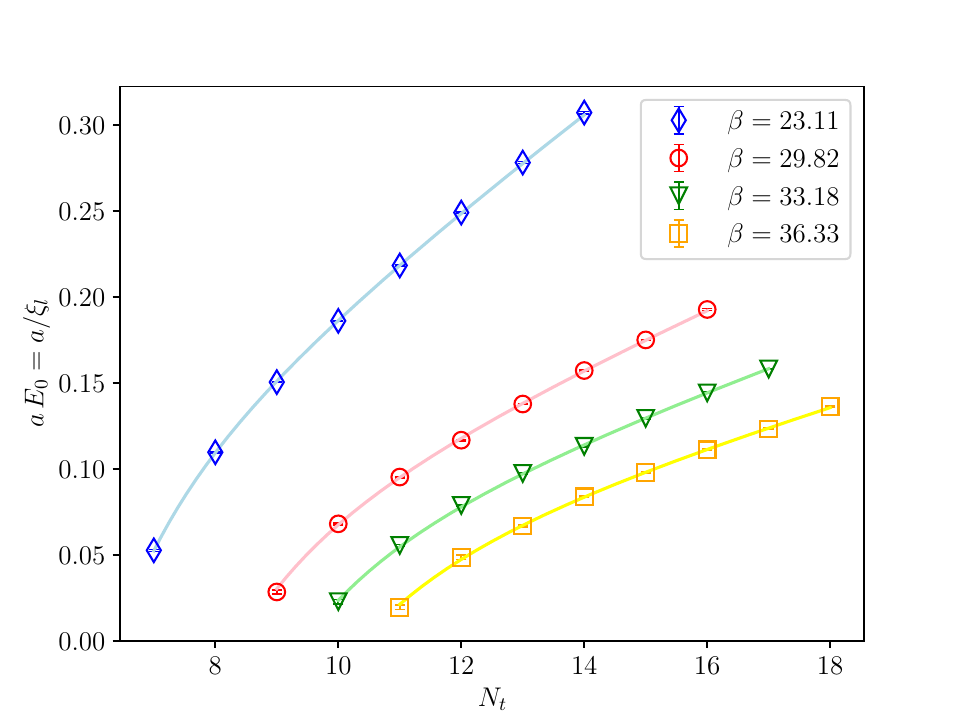}
 \caption{Combined best fits of the $\SU(3)$ ground-state energy $E_0$ for different values of $\beta$, according to eq.~\eqref{eq:BNG_groundstate_Nt11} including all terms up to $1/N_t^9$.}
 \label{fig:combined_fit_SU3_Nt9}
\end{figure}

\begin{figure}[H]
 \centering 
 \includegraphics[width=0.70\textwidth]{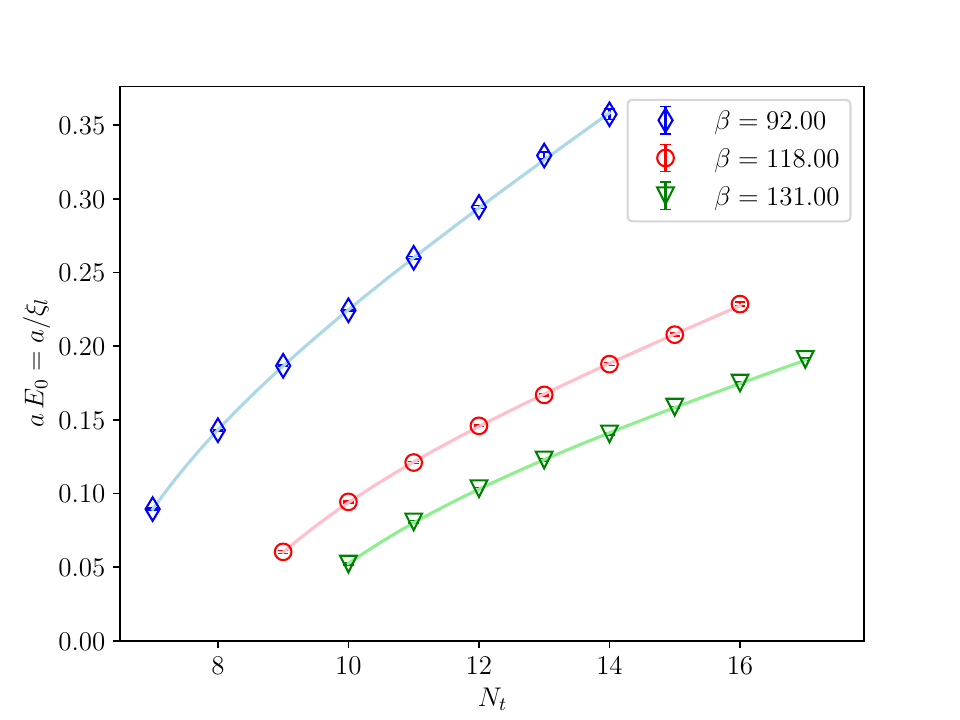}
 \caption{Combined best fits of the $\SU(6)$ ground-state energy $E_0$ for different values of $\beta$, according to eq.~\eqref{eq:BNG_groundstate_Nt11} including all terms up to $1/N_t^9$.}
 \label{fig:combined_fit_SU6_Nt9}
\end{figure}

\begin{figure}[H]
 \centering 
 \includegraphics[width=0.70\textwidth]{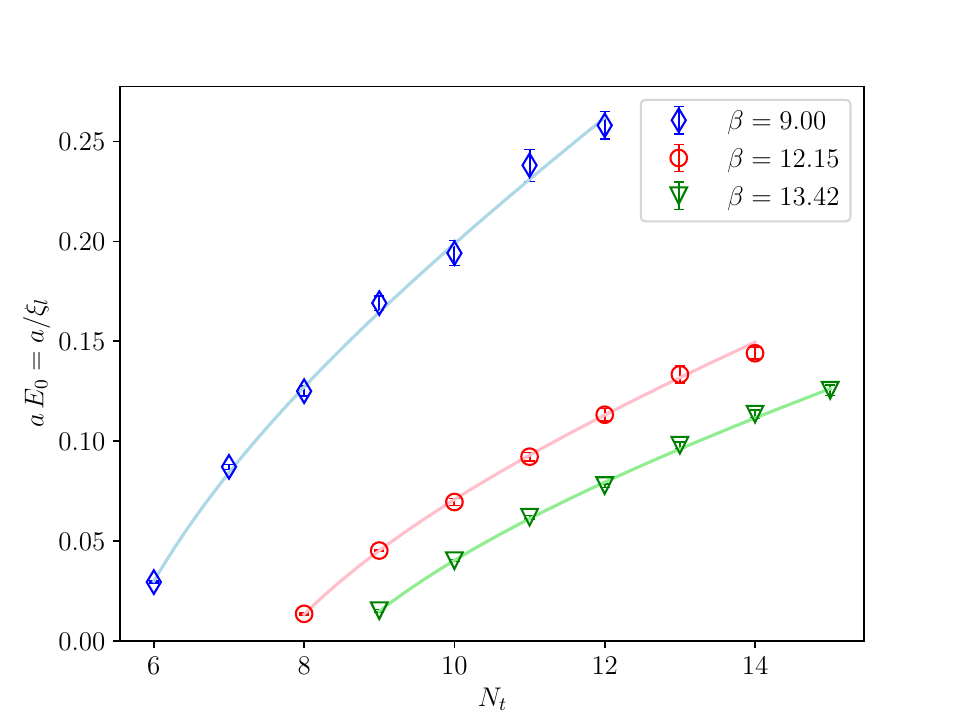}
 \caption{Combined best fits of the $\SU(2)$ ground-state energy $E_0$ for different values of $\beta$, according to eq.~\eqref{eq:BNG_groundstate_Nt11} including all terms up to $1/N_t^9$.}
 \label{fig:combined_fit_SU2_Nt9}
\end{figure}

\begin{table}[H]
\centering
\begin{tabular}{|l|l|l|l|l|l|l|l|l|}
\hline
$\beta$ & $N_t$ & & $R_{\mathrm{min}}/a$ & $R_{\mathrm{max}}/a$ & Amplitude & $\xi_l/a$ & $g_1$ & $\chi^2/N_{\mathrm{d.o.f.}}$ \\ \hline
\multirow{2}{*}{23.11} & \multirow{2}{*}{7} & $\eqref{eq:low_dist_corr_1}$ & 8 & 18 & $k_s=0.03099(13)$ & 19.07(15) & -1.558(2) & 1.6 \\ \cline{3-9} 
 & & $\eqref{eq:high_dist_corr}$ & 19 & 47 & $k_l=0.01831(16)$ & 19.00(21) & & 0.1 \\ \hline
\multirow{4}{*}{29.82} & \multirow{2}{*}{9} & $\eqref{eq:low_dist_corr_1}$ & 7 & 35 & $k_s=0.02726(11)$ & 31.5(4) & -1.549(3) & 1.10 \\ \cline{3-9} 
 & & $\eqref{eq:high_dist_corr}$ & 36 & 89 & $k_l=0.0142(8)$ & 35.6(1.5) & & 0.08 \\ \cline{2-9} 
 & \multirow{2}{*}{10} & $\eqref{eq:low_dist_corr_1}$ & 8 & 15 & $k_s=0.02794(15)$ & 15.31(12) & -1.5640(18) & 1.4 \\ \cline{3-9} 
 & & $\eqref{eq:high_dist_corr}$ & 15 & 79 & $k_l=0.01693(12$) & 14.70(12) & & 0.07 \\ \hline
\multirow{6}{*}{33.18} & \multirow{2}{*}{10} & $\eqref{eq:low_dist_corr_1}$ & 7 & 45 & $k_s=0.02587(13)$ & 38.6(5) & -1.539(5) & 0.6 \\ \cline{3-9} 
 & & $\eqref{eq:high_dist_corr}$ & 45 & 119 & $k_l=0.01320(10)$ & 44(3) & & 0.01 \\ \cline{2-9} 
 & \multirow{2}{*}{11} & $\eqref{eq:low_dist_corr_1}$ & 8 & 18 & $k_s=0.02688(13)$ & 18.16(15) & -1.560(2) & 1.9 \\ \cline{3-9} 
 & & $\eqref{eq:high_dist_corr}$ & 19 & 79 & $k_l=0.01582(17) $& 18.04(19) & & 0.05 \\ \cline{2-9} 
 & \multirow{2}{*}{12} & $\eqref{eq:low_dist_corr_1}$ & 8 & 13 & $k_s=0.02617(19)$ & 13.39(13) & -1.5661(19) & 0.9 \\ \cline{3-9} 
 & & $\eqref{eq:high_dist_corr}$ & 13 & 47 &$ k_l=0.01605(9) $ & 12.70(7) & & 0.05 \\ \hline
\multirow{8}{*}{36.33} & \multirow{2}{*}{11} & $\eqref{eq:low_dist_corr_1}$ & 8 & 52 & $k_s=0.02476(17)$ & 43.5(8) & -1.527(7) & 0.4 \\ \cline{3-9} 
 & & $\eqref{eq:high_dist_corr}$ & 53 & 119 & $k_l=0.0126(9)$ & 52(3) & & 0.03 \\ \cline{2-9} 
 & \multirow{2}{*}{12} & $\eqref{eq:low_dist_corr_1}$ & 9 & 21 & $k_s=0.02569(14)$ & 20.91(19) & -1.560(2) & 1.3 \\ \cline{3-9} 
 & & $\eqref{eq:high_dist_corr}$ & 21 & 79 & $k_l=0.0153(2) $& 20.6(5) & & 0.11 \\ \cline{2-9} 
 & \multirow{2}{*}{13} & $\eqref{eq:low_dist_corr_1}$ & 9 & 15 & $k_s=0.0253(2)$ & 15.3(2) & -1.559(3) & 1.6 \\ \cline{3-9} 
 & & $\eqref{eq:high_dist_corr}$ & 16 & 45 &$ k_l=0.01510(15) $ & 14.99(12) & & 0.4 
 \\ \cline{2-9} 
 & \multirow{2}{*}{14} & $\eqref{eq:low_dist_corr_1}$ & 9 & 13 & $k_s=0.0237(2)$ & 13.00(16) & -1.572(2) & 1.9 \\ \cline{3-9} 
 & & $\eqref{eq:high_dist_corr}$ & 13 & 47 & $k_l=0.01493(10) $& 11.94(6) & & 0.2 
 
\\ \hline 

\end{tabular}\\

\caption{Results of the fits to the short- and long-distance behavior, respectively encoded in eq.~\eqref{eq:low_dist_corr_1} and in eq.~\eqref{eq:high_dist_corr}, according to the Svetitsky--Yaffe mapping, of the Polyakov-loop correlator $G(R)$ for different values of $\beta$ and $N_t$.}
\label{tab:short_long}
\end{table}

\begin{table}[H]

\centering
\begin{tabular}{|c|c|c|c|c|c|c|}
\hline
 \multicolumn{7}{|c|}{$g_1 = -1.55938(85)$} \\
\hline
 $\beta$ & $N_t$ & $R_{\mathrm{min}}$ & $R_{\mathrm{max}}$ & $\xi_{short}$ & $\chi^2/N_{\mathrm{d.o.f.}}$ & $\xi_{long}$ \\
\hline
 23.11 & 7 & 8 & 18 & 19.14(12) & 1.5 & 19.00(21) \\
\hline
\multirow{2}{*}{29.82}
 & 9 & 7 & 35 & 32.02(30) & 1.4 & 35.6(1.5) \\
 & 10 & 8 & 15 & 15.050(86) & 2.4 & 14.70(12) \\
\hline
\multirow{2}{*}{33.18}
 & 10 & 7 & 45 & 39.77(48) & 0.96 & 44(3) \\
 & 11 & 8 & 18 & 18.11(11) & 1.7 & 18.04(19) \\
\hline
\multirow{3}{*}{36.33}
 & 11 & 8 & 52 & 45.65(78) & 0.70 & 52(3) \\
 & 12 & 9 & 21 & 20.85(13) & 1.0 & 20.6(5) \\
 & 13 & 9 & 15 & 15.290(92) & 1.6 & 14.99(12) \\
\hline
\end{tabular}
\caption{Same as in table~\ref{tab:short_long}, but performing a combined fit of data for $G(R)$ to eqs.~\eqref{eq:low_dist_corr_1} and~\eqref{eq:high_dist_corr} for different values of $\beta$ and $N_t$.}
\label{tab:short_long2}
\end{table}

\end{appendices}

\bibliographystyle{JHEP}
\bibliography{beyondNG.bib}

\end{document}